\documentclass[twocolumn,showpacs,preprintnumbers,prd,nofootinbib]{revtex4-1}
\usepackage{feynman}
\usepackage{graphicx,amsmath,amssymb,xcolor,hyperref}

\begin{document}

\author{Yi Chung}
\email{yi.chung@mpi-hd.mpg.de}
\author{Florian Goertz}
\email{florian.goertz@mpi-hd.mpg.de}
\affiliation{Max-Planck-Institut f\"ur Kernphysik, Saupfercheckweg 1, 69117 Heidelberg, Germany}

\title{Third-generation-philic Hidden Naturalness}

\begin{abstract}
We present a solution to the electroweak hierarchy problem, where the relevant new particles are third-generation-philic and hidden in SM processes with third-generation fermions. Due to this feature, the mass bounds from direct searches are much weaker and the required fine-tuning can be reduced drastically.
A concrete model is constructed based on a $SU(6)/Sp(6)$ fundamental composite Higgs model with collective symmetry breaking and extended hypercolor mechanism. The construction allows to raise the scale $f$ to $\sim 3\,$TeV, corresponding to resonances at $M_\rho \gtrsim 10$\, TeV, without much tuning -- employing ingredients that are naturally inherent in the (composite) Goldstone-Higgs framework. The experimental signatures are discussed in detail. It is found that current bounds allow for a model with negligible tuning.
\end{abstract}

\maketitle

\section{Introduction}

The Standard Model (SM) of particle physics describes all known particles and interactions successfully. At the center of the SM is the mechanism of electroweak symmetry breaking (EWSB), which is responsible for the masses of gauge bosons and fermions. The discovery of the Higgs bosons in 2012 \cite{Chatrchyan:2012xdj, Aad:2012tfa} filled in the last missing piece of the SM. However, the SM does not address the UV-sensitive nature of the Higgs field. Its mass term receives quadratically divergent radiative corrections from its interactions with other SM fields and also itself, which leads to the well-known hierarchy problem. The three main corrections come from the top quark loop, weak boson loop, and Higgs boson loop, reading
\begin{align}\label{deltam}
\delta m_H^2= 
-\frac{3}{8\pi^2}y_t^2\Lambda_t^2+\frac{9}{64\pi^2}g_W^2\Lambda_g^2+\frac{3}{8\pi^2}\lambda\Lambda_h^2~.
\end{align}
Following the Naturalness principle, we expect new degrees of freedom to show up at the scales $\Lambda_t\sim 500 \text{ GeV}$, $\Lambda_g\sim 1200 \text{ GeV}$, and $\Lambda_h\sim 1300 \text{ GeV}$, respectively.

However, because of the absence of new physics in LHC searches, the bounds on the mass of expected new particles in traditional models, such as supersymmetry (SUSY) and composite Higgs model (CHM), have already surpassed the values shown above and challenge the Naturalness principle due to the growing required tuning. To avoid fine-tuning, one alternative is to have new particles being SM singlets which could be significantly lighter while escaping direct detection at the LHC. The idea is known as Neutral Naturalness, realized for example in twin Higgs models \cite{Chacko:2005pe}.

In this letter, we present a new solution where the new particles have a different nature and can still be detected directly at the LHC, but only or mainly through third-generation-fermion channels. Such a feature can significantly reduce the production rate and make the final states harder to measure at the same time, which might explain why they have not been observed by the LHC so far. Moreover, these third-generation-philic particles also have weaker constraints from flavor observables due to an accidental $U(2)^5$ flavor symmetry \cite{Barbieri:2011ci,Isidori:2012ts}.

\section{A Concrete Model}\label{sec:Model}

A concrete model avoiding tuning is constructed, starting from a fundamental composite Higgs model (FCHM)\cite{Cacciapaglia:2014uja, Cacciapaglia:2020kgq}, by employing (1) \textbf{Collective Symmetry Breaking} from little Higgs models \cite{ArkaniHamed:2001nc, ArkaniHamed:2002qx, ArkaniHamed:2002qy, Low:2002ws} with heavy $SU(2)'$ bosons to cancel the SM weak boson loops, which are further introduced to be third-generation-philic, and (2) \textbf{Extended Hypercolor (EHC)} \cite{Barnard:2013zea, Ferretti:2013kya, Cacciapaglia:2015yra} with low-scale top-philic EHC bosons to cut off the top loop, based on the idea presented in \cite{Bally:2023lji,Chung:2023iwj}, and more generally in~\cite{Bally:2022naz}.

To obtain the desired third-generation-philic feature, we start with extending the $SU(2)_W$ gauge group of the SM to $SU(2)_I$ and $SU(2)_{I\!I}$. The third generation fermions are charged under both $SU(2)$, as will be explained in Sec. \ref{sec:Top}, and the light generations are only charged under $SU(2)_I$, similar to \cite{Malkawi:1996fs, Muller:1996dj, Chiang:2009kb}. 

Next, we introduce the hypercolor group $\mathcal{G}_{HC}$ and hyperfermions $\psi$ which constitute the strong sector of our FCHM. The hyperfermions are representations of the hypercolor group $\mathcal{G}_{HC}$, whose coupling becomes strong around the TeV scale. The hyperfermions then form a condensate, which breaks the global symmetry and realizes the Higgs doublet as a set of pseudo-Nambu-Goldstone bosons (pNGBs).  To get the desired properties, we need six Weyl hyperfermions transforming under $SU(2)_I\times SU(2)_{I\!I} \times U(1)_Y$ as
\begin{align}
Q_{I,L}=(2,1)_0,~ Q_{I\!I,L}=(1,2)_0,~ T_R^c/B_R^c=(1,1)_{\mp\frac{1}{2}}.
\end{align}
Assuming that they transform according to the same pseudo-real representation of the hypercolor group $\mathcal{G}_{HC}$, they can be combined as
\begin{equation}
\psi=(Q_{I,L},T_R^c,B_R^c, (-i\sigma^2) {Q}_{I\!I,L})^T
\end{equation}
with $\sigma^2$ acting on the $SU(2)_{I\!I}$ index. The hyperfermion $\psi$ exhibits a global $SU(6)$ symmetry (partially gauged). The condensate $\psi\psi$ breaks $SU(6)$ to its $Sp(6)$ subgroup, which contains the breaking $SU(2)_I\times SU(2)_{I\!I} \to SU(2)_W$. The $SU(6)/Sp(6)$ coset is one of the earliest cosets employed in little Higgs models \cite{Low:2002ws} with collective symmetry breaking and is also considered in refs. \cite{Katz:2005au, Cai:2019cow,Rosenlyst:2020znn,Cheng:2020dum,Cacciapaglia:2020psm,Chung:2021fpc,Rosenlyst:2021tdr} for different prospects. 


\subsection{Basics of $SU(6)/Sp(6)$}

To study the $SU(6)/Sp(6)$ symmetry breaking, we can parametrize it by a non-linear sigma model. Consider a scalar field $\Sigma$, which transforms as an anti-symmetric tensor representation $\mathbf{15}$ of $SU(6)$. The transformation can be expressed as $\Sigma \to g\,\Sigma \,g^T$ with $g\in SU(6)$. The field $\Sigma$ has an anti-symmetric vacuum expectation value (vev) given by ($\epsilon=i\sigma^2$)
\begin{equation}
\langle \Sigma\rangle = \Sigma_0=
\begin{pmatrix}
0  &  0  &  -\mathbb{I}_{2\times 2} \\
0  &  \epsilon  &  0 \\
\mathbb{I}_{2\times 2}   &  0  &  0 \\
\end{pmatrix}\,,
\end{equation}
which breaks $SU(6)$ down to $Sp(6)$ and produces a total of 14 NGBs. 

The $SU(6)$ generators are divided into 
\begin{equation}
\begin{cases}
\text{unbroken generators}   &~T_a   : T_a\Sigma_0+\Sigma_0T_a^T=0~,\\
\text{broken generators}       &X_a   : X_a\Sigma_0-\Sigma_0X_a^T=0~.
\end{cases}
\end{equation}
The NGBs can be written as a matrix with the broken generators (and symmetry-breaking scale $f$)
\begin{equation}
\xi(x)\equiv e^{\frac{i\pi_a(x)X_a}{2f}}\,,
\end{equation}
transforming under $SU(6)$ as $\xi \to g\, \xi \,h^{\dagger}$ with $h \in Sp(6)$ and
\begin{equation}
\Sigma(x)= \xi(x)\, \Sigma_0\,\xi^T(x)=e^{\frac{i\pi_a(x)X_a}{f}}\Sigma_0~.
\end{equation}
They are structured as
\begin{align}
i\pi_aX_a \!\cdot\! \Sigma_0\!=\!
\begin{pmatrix}
\epsilon s  & \left({H}_1\,\tilde{H}_2\right)   &     -\frac{\phi_a}{\sqrt{2}}\sigma^a\!+\!\frac{\eta}{\sqrt{6}}\mathbb{I}_{}        \\
-\left({H}_1\,\tilde{H}_2\right)^T &  -2\frac{\eta}{\sqrt{6}}\epsilon  & \left(\tilde{H}_2\,H_1\right)^\dagger   \\
 \frac{\phi_a}{\sqrt{2}}\sigma^{a*}\!\!-\!\frac{\eta}{\sqrt{6}}\mathbb{I}_{} &   \left(-\tilde{H}_2\,H_1\right)^{\!*} &  - \epsilon^T s^*   \\
\end{pmatrix}\!,
\end{align}
forming a real $SU(2)_W$ triplet $\phi_a$ (which will be absorbed by the massive $SU(2)'$ gauge bosons), a complex singlet $s$, a real singlet $\eta$, and two complex Higgs doublets $H_1$ and $H_2$, both with hypercharge $-1/2$. We will show in the next section that the latter feature a two-Higgs-doublet model (2HDM) with tan$\,\beta=1$ and can naturally realize the SM alignment limit $\beta-\alpha=\pi/2$, resulting in an inert second Higgs doublet. The observed Higgs boson $h$ is a linear combination $(h_1+h_2)/\sqrt{2}$, where $h_1$ and $h_2$ are the neutral CP-even components of the two Higgs doublets. 


\subsection{The gauge sector}\label{sec:Gauge}

The gauge group $SU(2)_I\times SU(2)_{I\!I}$ with couplings $g_I$ and $g_{I\!I}$ is embedded in $SU(6)$ with generators
\begin{align}
\frac{1}{2}
\begin{pmatrix}
\sigma^a    &  0  &  0 \\
0   &  0  &  0  \\
0   &  0  &  0 \\
\end{pmatrix} \quad\text{and}\quad
\frac{1}{2}&
\begin{pmatrix}
0   &  0  &  0  \\
0   &  0  &  0 \\
0   &  0  &  -\sigma^{a*} \\
\end{pmatrix}~.\nonumber
\end{align}
The vev $\Sigma_0$ directly breaks $SU(2)_I\times SU(2)_{I\!I}$ down to the SM $SU(2)_W$ gauge group with coupling given by $g_W=g_I g_{I\!I} / \sqrt{g_I^2+g_{I\!I}^2}$. The broken $SU(2)'$ gauge bosons get masses
\begin{equation}
M_{W'}^2 \approx M_{Z'}^2 \approx \frac{1}{4}\,g_{W'}^2f^2,
\end{equation}
where $g_{W'}=\sqrt{g_I^2+g_{I\!I}^2}$. The SM group $SU(2)_W\times U(1)_Y$ is embedded in the unbroken $Sp(6)$ with generators
\begin{equation}
\frac{1}{2}
\begin{pmatrix}
\sigma^a    &  0  &  0 \\
0   &  0  &  0  \\
0   &  0  &  -\sigma^{a*} \\
\end{pmatrix}\quad\text{and}\quad
\frac{1}{2}
\begin{pmatrix}
0   &  0  &  0 \\
0   &  -\sigma_3  &  0  \\
0   &  0  &  0 \\
\end{pmatrix} ~.
\end{equation}

The kinetic Lagrangian for the NGBs is given by
\begin{equation}
\mathcal{L}=\frac{f^2}{8}\text{tr}\left[(D_{\mu}\Sigma)(D^\mu \Sigma)^\dagger\right] ,
\label{Lagrangian}
\end{equation}
where $D_{\mu}$ is the electroweak covariant derivative. After expanding, the Lagrangian for the Higgs boson $h$ reads
\begin{equation}
\mathcal{L}_h=\frac{1}{2}(\partial _\mu h)^2+\frac{f^2}{8}g_W^2 \,\text{sin}^2\!\left(\frac{h}{f}\right) \left[2W^+_\mu W^{-\mu}+\frac{Z_\mu Z^\mu}{\text{cos}^2\,\theta_W}\right]\!,
\end{equation}
with the non-linear behavior of the CHM being apparent in the trigonometric functions. When $h$ obtains a vev $\langle h\rangle=V$, the weak bosons acquires a mass of
\begin{equation}
m_W^2={m_Z^2}{\,\text{cos}^2\,\theta_W}=\frac{f^2}{4}g_W^2 \,\text{sin}^2\left(\frac{V}{f}\right)=\frac{1}{4}\,g_W^2v^2,
\end{equation}
where $v\equiv f\,\text{sin}\,(V/f)\approx V$.
The Higgs non-linearity
\begin{equation}
\xi\equiv \frac{v^2}{f^2}= \sin^2\left(\frac{V}{f}\right)
\end{equation}
parametrizes deviations of the Higgs couplings from the SM, which for vector bosons become
\begin{equation}
\kappa_V\equiv \frac{g_{hVV}}{g^{SM}_{hVV}}=
\text{cos}\left(\frac{V}{f}\right)=\sqrt{1-\xi}\approx 1-\frac{\xi}{2}~.
\end{equation}

\subsection{The top sector}\label{sec:Top}

In CHMs, the top Yukawa coupling arises from a higher dimensional operator. The scale of the operator $\Lambda_{t}$ labels its origin, which plays the role of the cutoff of the top loop in Eq.~\eqref{deltam}. For example, in our FCHM setup, the top Yukawa arises from $D\!=\!6$ four-Fermi operators\footnote{While we use the bilinear term for simplicity, the idea can also be realized with linear coupling, i.e.\ partial compositeness~\cite{Kaplan:1991dc}.}
\begin{align}\label{yteff}
\mathcal{L}_{\text{}}=\sum_{i=I,I\!I}\, \frac{g_t^2}{\Lambda_{t}^2}
\left({\bar{T}_R}{Q}_{i,L}\right)
\left({\bar{q}_{i,L}}{t}_{R}\right)~,
\end{align}
where the third-generation left-handed SM quark corresponds to the combination $q_L =  (q_{I,L}+q_{I\!I,L})/\sqrt2$
of $SU(2)_I$ and $SU(2)_{I\!I}$ doublet quarks to achieve a viable phenomenology.
Once hypercolor becomes strongly coupled, ${\bar{T}_R}\,{Q}_{I\!/\!I\!I,L}$ form bound states $H_{1/2}$ and the top Yukawa coupling is generated. Such an effective operator can arise from FCHM with an extended gauge group, where a new gauge group $\mathcal{G}_{E}$ is introduced to connect the top quarks and hyperfermions.\footnote{We assume $\mathcal{G}_{E}$ specifically generates the top Yukawa and therefore is top-philic, neglecting the other Yukawa couplings. It would be interesting to explicitly construct a corresponding model where the light generations are naturally realized via partial compositeness, which is indeed somewhat problematic only for the heavy top quark.} The relevant Lagrangian is given by
\begin{align}\label{LEHC}
{\cal L}_{\text{E}}=\sum_{i=I,I\!I} g_{E} {E}_{\mu} (\bar{Q}_{i,L}\gamma^\mu q_{i,L}+\bar{T}_R\gamma^\mu t_R)~.
\end{align}
Integrating out the massive $E_\mu$ gauge bosons, the term in Eq.~\eqref{yteff} is reproduced with $g_t\to g_{E}$ and $\Lambda_{t} \to M_{E}$. 

We finally obtain the Yukawa term
\begin{align}\label{topyukawa}
\mathcal{L}={y_t}\left({\bar{q}_{I,L}}H_1\,{t}_{R}+{\bar{q}_{I\!I,L}}H_2\,{t}_{R}\right)\supset \frac{y_t}{\sqrt{2}}\,\bar{q}_{L}\left(H_1+H_2\right)\,{t}_{R},
\end{align} after formation of the hyperfermion condensate, where
\begin{align}\label{topyukawa2}
y_t\sim \frac{1}{v}\frac{g_{E}^2}{M_{E}^2}\langle \bar{T}_R{Q}_{I\!/\!I\!I,L}\rangle_{HC}
\sim Y_s \left(\frac{f}{f_{E}}\right)^2
\end{align}
is universal since stemming from a single EHC mechanism and we neglected the trigonometric factor due to the Higgs non-linearity for simplicity.
Above, $Y_s$ is the Yuakwa coupling from the strong dynamics, such as $Y_s \sim 3$ in QCD, and $f_{E} \equiv M_{E}/g_{E} = \Lambda_{t}/g_t$ represents the breaking scale of $\mathcal{G}_{E}$.
Therefore, the third generation fermions couple to both $SU(2)$'s and $SU(2)_{I\!I}$ is third-generation-philic.
The top Yukawa coupling arising from this mechanism will share the same deviation due to the non-linearity of the pNGB Higgs as vector boson couplings, i.e. $\kappa_t=\kappa_V$. The recent Higgs fits require $\xi \lesssim 0.1$ \cite{Khosa:2021wsu}, implying $f \gtrsim 0.8$ TeV, which will be considered in the following analysis.

\section{The pNGB potential} \label{sec:Potential}

The interaction terms introduced in the last section break the $SU(6)$ global symmetry explicitly and thereby will contribute to the potential of the pNGB fields. Starting with the gauge interactions, $SU(2)_I$ and $SU(2)_{I\!I}$ introduce the potential terms
\begin{align}
c\,g_I^2f^2\left|s+\frac{i}{2f}{H_2}^\dagger H_1\right|^2+
c\,g_{I\!I}^2f^2\left|s-\frac{i}{2f}{H_2}^\dagger H_1\right|^2,
\label{collective_term}
\end{align}
where $c \sim {\cal O}(1)$ is determined by the concrete UV model. The terms break the shift symmetry of the complex scalar $s$ and induce the mass term
\begin{align}
V_s=c\,(g_I^2+g_{I\!I}^2)f^2s^2=c\,g_{W'}^2f^2s^2~.
\end{align} 
Besides, there are also trilinear couplings between $s$ and the Higgs fields given by
\begin{align}
V_{sHH}=\frac{i}{2}cf\,(g_I^2-g_{I\!I}^2)\,s{H_2}^\dagger H_1\,  + {\rm h.c.}~.
\label{sHH}
\end{align}
A nontrivial Higgs quartic term is also generated after combining terms in Eq. \eqref{collective_term} and the tree-level contribution from integrating out $s$ as
\begin{align}
V_{\text{quartic}}=c\,\frac{g_I^2g_{I\!I}^2}{g_I^2+g_{I\!I}^2}\left|{H_2}^\dagger H_1\right|^2=c\,g_W^2\left|{H_2}^\dagger H_1\right|^2~,
\end{align}
which is the key feature of little Higgs models.

A loop level potential for $H_1$ and $H_2$ is induced from the gauge bosons as
\begin{equation}
\begin{split}
V_{g}=\frac{9}{64\pi^2}g_W^2\,\text{ln}\frac{\Lambda^2}{M_{W'}^2} & \Big{[}M_{W'}^2\left(|H_1|^2+|H_2|^2\right)\nonumber  \\ & +\frac{g_{W'}^2}{8}\left(|H_1|^2+|H_2|^2\right)^2\Big{]}\,,
\end{split}
\end{equation}
where $\Lambda \sim 4\pi f$ is the cutoff of the theory. Contributions from other gauge interactions, such as $U(1)_Y$ (and $U(1)_{H\!B}$, introduced below), are suppressed due to weaker couplings and not discussed here.

Finally, the top Yukawa in Eq. \eqref{topyukawa} breaks the shift symmetry of $H_1+H_2$, leading to
\begin{align}
V_{t}=-\frac{3}{16\pi^2}y_t^2\left( \Lambda_{t}^2|H_1+H_2|^2 \!- \frac{y_t^2}{4} \text{ln}\frac{\Lambda_{t}^2}{m_{t}^2} |H_1+H_2|^4 \right)\,,
\end{align}
where $\Lambda_{t}$ is the cutoff of the top loop contribution. In CHMs with EHC, this corresponds to the $E_\mu$ boson mass $M_{E}=g_{E}f_{E}$, which can be light if $g_{E}$ is weak.

\section{The Higgs potential in the 2HDM and Naturalness}\label{sec:VH}

The doublet part of the potential corresponds to a 2HDM with 
\begin{align}
V_H&= V_t+V_g +V_\text{quartic}~,
\end{align}
featuring a $S_2$ symmetry between $H_1$ and $H_2$, which guarantees $\langle H_1\rangle = \langle H_2\rangle$. A natural alignment limit with tan$\beta$=1 is automatically satisfied. In the Higgs basis
\begin{align}
H_\text{SM}=\frac{1}{\sqrt{2}}(H_1+H_2),\quad
H_\text{inert}=\frac{1}{\sqrt{2}}(H_1-H_2)\,,
\end{align}
where $\langle H_\text{SM}\rangle =v/\sqrt{2}$ and $\langle H_\text{inert}\rangle =0$,
the potential reads
\begin{align}
V_H
&=(m_{H,t}^2+m_{H,g}^2)|H_\text{SM}|^2+(\lambda_t+\lambda_g+\lambda_s)|H_\text{SM}|^4\nonumber\\
&+(m_{H,g}^2)|H_\text{inert}|^2+(\lambda_g+\lambda_s)|H_\text{inert}|^4\nonumber\\
&+(2\lambda_g-\frac{2\lambda_s}{3} )|H_\text{SM}|^2|H_\text{inert}|^2\nonumber\\
&+\frac{2\lambda_s}{3} \left(|H_\text{SM}^\dagger H_\text{inert}|^2-\frac 1 2 \left[(H_\text{SM}^\dagger H_\text{inert})^2 + {\rm h.c.}\right]\right)~,
\end{align}
with 
\begin{align}\label{Vcoeff}
&m_{H,t}^2=-\frac{3}{8\pi^2}y_t^2M_{E}^2\,,~
m_{H,g}^2=\frac{9}{64\pi^2}g_W^2M_{W'}^2\,\text{ln}\frac{\Lambda^2}{M_{W'}^2}~,\nonumber\\
&\lambda_t =\frac{3y_t^4}{16\pi^2}\text{ln}\frac{M_{E}^2}{m_{t}^2},~ \lambda_g=\frac{9g_W^2g_{W'}^2}{512\pi^2}\text{ln}\frac{\Lambda^2}{M_{W'}^2},~
\lambda_s =\frac{c\,g_W^2}{4}.\nonumber
\end{align}

The $H_\text{SM}$ part should match the SM Higgs potential
\begin{align}
V_\text{SM}=-\mu^2_\text{SM}|H_\text{SM}|^2+\lambda_\text{SM}|H_\text{SM}|^4~,
\end{align}
where the coefficient of the quadratic term (relevant for Naturalness) is given by 
\begin{align}
-\mu^2_\text{SM}&=-\frac{3}{8\pi^2}y_t^2M_{E}^2+\frac{9}{64\pi^2}g_W^2M_{W'}^2\,\text{ln}\frac{\Lambda^2}{M_{W'}^2}~,
\end{align}
in line with the IR potential from Eq.~\eqref{deltam} with top loop and gauge loop contribution.

The former is cut off by $M_{E}=g_Ef_E$, which now replaces $f$ as the fundamental scale and can be significantly smaller for small $g_{E}$. The Naturalness principle requires $M_{E}\sim 500$\,GeV. Notice that the top loop contribution is now totally independent of the scale $f$.

The mass of the heavy boson entering the gauge loop is given by $M_{W'}=1/2\, g_{W'}f$. Different from conventional CHMs, where the loop is cut off by the vector resonances of the strong sector with $M_\rho \sim g_\rho f$, now it is canceled by the heavy $SU(2)'$ boson in the little Higgs setup. Compared to the large $g_\rho$ from the strong dynamics, $g_{W'}$ can be smaller and thus a lighter $M_{W'}$ can lower down $m_H^2$ and the fine-tuning. Numerically, Naturalness requires $M_{W'}\sim 1.2$ TeV, which restricts the combination $g_{W'} f$. If some level of fine-tuning, such as $25\% \,(10\%)$, is allowed, we can obtain a heavier $M_{W'}\sim 2.3\,(3.6)$ TeV.

Most importantly, the new particles at the cutoff of both loops are third-generation-philic, which means their direct search bound is much weaker compared to other scenarios: The top loop is cut off by the EHC boson, which can couple only to top quarks, and the $W'$ boson, being a linear combination of $W_I$ and $W_{I\!I}$, is also third-generation-philic if $g_{I\!I} \gg g_I$. More phenomenology will be discussed in the next section.

The Higgs quartic emerges from three different sources,
\begin{align}
\lambda_\text{SM}&=\lambda_s+\lambda_t+\lambda_g=
\frac{c\,g_W^2}{4}+(\text{one-loop})~,
\end{align}
but is dominated by the tree-level contribution as in little-Higgs type models. Some contributions from the top loop and gauge loop are also required to reach the observed value of $\lambda_\text{SM}=0.13$.

Considering the dominant contributions, a natural splitting between $v$ and $f$ is expected as
\begin{align}
\frac{v}{f} = \frac{1}{f}\sqrt{\frac{-m_{H,t}^2}{\lambda_s}} \sim \sqrt{\frac{3Y_s y_t}{2\pi^2}}\frac{g_{E}}{g_W}~,
\end{align}
which becomes $\sim 1/3$ for $Y_s=3$ and $g_W=2g_{E}$.
Therefore, Naturalness is realized two-fold: First, the quadratic term from the UV theory can still be naturally small thanks to the weaker constraints for third-generation fermions, and second, $v/f$ is naturally small thanks to collective symmetry breaking, which solves the little hierarchy problem. The residual tuning is fully captured by the ratio of the numerator above to its natural value $\sqrt{- \mu^2_\text{SM}} \sim 100\,$GeV, as discussed before.
We note that for very small $g_E$ a huge separation of scales seems possible in a natural way, which is however limited by (above-neglected) gauge contributions $m_{H,g}^2 \propto g_{W'}^2$, which are bounded from below by direct searches and by $g_{W'} > 2 g_W$, see Fig.~\ref{taunu}.


So far, we have neglected the quadratic term arising from Higgs self-interactions as shown in Eq.~\eqref{deltam}, which receives different contributions with different cutoffs,
\begin{align}
\delta m_{H,\lambda}^2= \frac{3}{8\pi^2}
\left(\lambda_s M_{s}^2+\lambda_t M_{E}^2+\lambda_g M_{W'}^2\right)~,
\end{align}
where we ignore ${\cal O}(1)$ logarithms. The second and third terms are two-loop suppressed and thus negligible, whereas the first term leads to
\begin{align}
\delta m_{H,\lambda_s}^2= \frac{3}{32\pi^2}
c^2g_W^2g_{W'}^2f^2~,
\end{align}
which exhibits a similar form as the gauge contribution. Having the Higgs-loop contribution smaller requires $c\lesssim\sqrt{3/8}\sim 0.6$, leading to $\lambda_s \lesssim 1/2\, \lambda_\text{SM}$ and $M_s\lesssim 1.5 M_{W'}$.

\section{The spectrum of new particles}

We recall that the masses of the new gauge bosons from the extended gauge group, including $SU(2)'$ and EHC, and of the complex scalar singlet, which are all crucial to cut off the different loop corrections to the Higgs mass, are given by
\begin{equation}
\begin{split}
M_{W'} \approx & M_{Z'} \approx \frac{1}{2}\,g_{W'}f~,\\
M_{E} = & g_{E}f_E \approx g_{E} \sqrt{Y_s/y_t} \, f~,\\
M_s= & \sqrt{c}\,g_{W'}f~.
\end{split}
\end{equation}

Moreover, the masses of the inert doublet
\begin{equation}
H_{\rm inert} = \begin{pmatrix}
H^{+} \\ 
\left( H + iA \right)/\sqrt{2}
\end{pmatrix}
\end{equation}
read
\begin{align}
M_H^2= M_{H^\pm}^2=m_{H,g}^2+(\lambda_{g}-\frac{\lambda_s}{3})\,v^2\,,\\ \nonumber
M_A^2=m_{H,g}^2+(\lambda_{g}+\frac{\lambda_s}{3})\,v^2 ~,
\end{align} 
clustered around $ m_{H,g}$. Since the latter also enters the SM Higss potential, the mass of the inert doublet can range from $\sim 100$ GeV (without fine-tuning) to $\sim 300$ GeV ($10\%$ fine-tuning).

Finally, the real singlet $\eta$ remains massless so far. 
To address this, further explicit breaking via a hyperfermion mass term could be introduced. In the following, we will however avoid the massless singlet by gauging the corresponding $U(1)_{H\!B}$ symmetry, where $H\!B$ denotes hyperbaryon number~\cite{Chung:2021ekz,Chung:2021xhd}, which is broken by $\Sigma_0$. In this case, we obtain a neutral, TeV-scale boson $Z'_{H\!B}$,
\begin{align}
M_{Z'_{H\!B}}\approx \frac{2}{N_{HC}}g_{H\!B}f~,
\end{align} 
where $N_{HC}$ is the number of hypercolors.


\section{Collider Searches}\label{sec:Collider}

In this section, we explore collider searches for the new states in dependence on the cutoff scales.

\subsection{Cutoff of gauge loop : Heavy gauge boson}

\begin{figure}[tbp]
\centering
\includegraphics[width=0.48\textwidth]{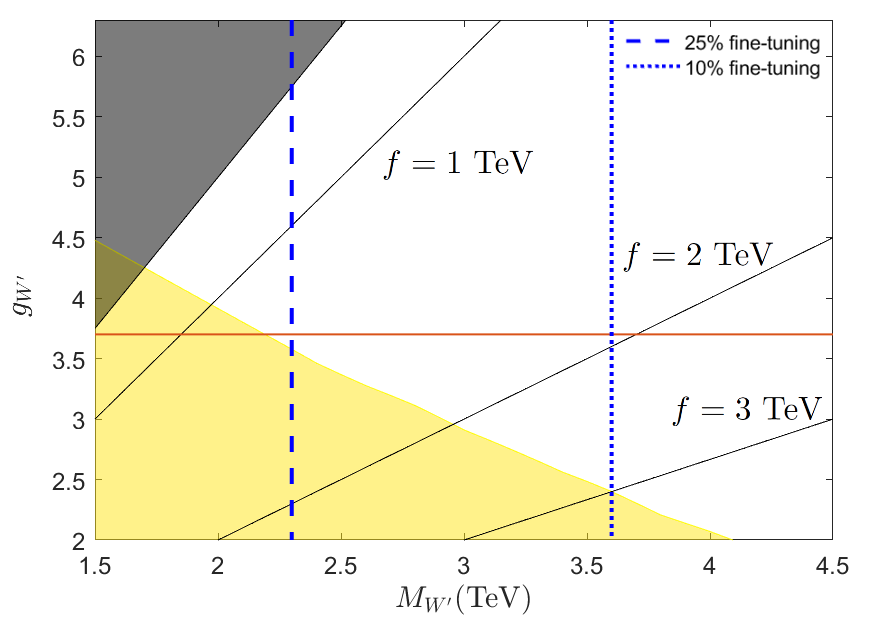}
\caption{Constraints on $g_{W'}$ v.s. $M_{W'}$. The yellow region is excluded by the CMS 138 fb$^{-1}$ $\tau\nu$ direct search. Above the brown line, $\Gamma_{W'}/M_{W'}>30\%$ and the direct search is no longer valid. The three black lines represent different values of $f$, and the black region is excluded by Higgs coupling measurements. Two blue lines correspond to different levels of fine-tuning.  \label{taunu}}
\end{figure}

The phenomenology of the heavy $SU(2)'$ bosons is similar to the 221 nonuniversal gauge interaction model \cite{Malkawi:1996fs, Muller:1996dj, Chiang:2009kb}, with the mixing angle cot$\,\theta_e=g_{I\!I}/g_I$. 
The heavy $W'$ boson corresponds to the linear combination sin$\,\theta_eW_I+$cos$\,\theta_eW_{I\!I}$. For large cot$\,\theta_e$, i.e. $g_{I\!I} \gg g_I$, sin$\,\theta_e$ is suppressed and $W'$ becomes third-generation-philic, with the main constraints originating from final states with third generation leptons.

For the charged boson, the strongest bound comes from $W'\to \tau\nu$ \cite{CMS:2022ncp, ATLAS:2021bjk}, while  $W'\to tb$ \cite{CMS:2023ldh,ATLAS:2021drn} delivers a weaker constraint. The exclusion plot in the $g_{W'}$--$M_{W'}$ plane is shown in Fig.~\ref{taunu} with the constraint from Higgs-coupling measurements discussed in Sec. \ref{sec:Top}, while indicating the level of fine-tuning by blue-dashed lines.

For $25\%$ fine-tuning or better, we obtain a rather small triangle in the left, where most of the parameter space features a rather large $g_{W'} \lesssim5.7$ corresponding to a broad $W'$ resonance, which requires new search methods. There is also an upper bound on the scale $f<1.3$ TeV from the bottom of the triangle. We note that in the left corner, at $M_{W'} \sim 1.7\,$TeV, there is basically no tuning. Allowing up to $10\%$ fine-tuning, a larger parameter space with $g_{W'}$ as small as $2.4$ and a larger $f=3$ TeV is allowed. Note that, in conventional CH models, this would already correspond to tuning exceeding $1\%$, more than a factor of 10 worse. The upper edge is bounded by $g_{W'}<6.3$, above which the model becomes nonperturbative and bound states will form. 

For the neutral boson ($Z^\prime \equiv W^\prime_3$), the constraints are in general milder with the main one originating from the $Z'\to \tau\tau$ decay~\cite{CMS:2022goy,ATLAS:2020zms}, while $Z'\to t\bar{t}$ \cite{ATLAS:2020lks} and $Z'\to b\bar{b}$ \cite{CMS:2022zoc,ATLAS:2019fgd} feature even weaker bounds.

\subsection{Cutoff of top loop : EHC gauge boson}

The top Yukawa coupling is generated through the interaction in Eq.~\eqref{LEHC}, where the detailed properties of the EHC gauge boson depend on the complete extended hypercolor model, see e.g. \cite{Chung:2023iwj}. However, searches for a $E_\mu$ resonance are in general not very promising as it should be lighter than the hypercolor states for models with little fine-tuning, which will forbid common two-body decays. Instead, it can only be searched through loop effects hidden in SM processes. For example, if $E_\mu$ is colored, an operator $\mathcal{O}_{tG}=g_s\left(\bar{q}_L\sigma^{\mu\nu}T_At_R\right)\tilde{H}G^A_{\mu\nu}+\text{h.c.}$ will be generated through a loop with $E_\mu$ bosons and a hyperfermion and affects the $\bar{t}t$ final state \cite{CMS:2019nrx}. The loop suppression makes the current constraint rather weak but the desired precision can be reached in the near future. Besides, we also expect contributions to other final states as $t\bar{t}h\,/\,t\bar{t}Z\,/\,tbW$, which can be tested at the HL-LHC.

Moreover, in most UV models, a neutral massive $Z'_{E}$ boson from the diagonal $U(1)$ of the EHC group appears. In our case, both $q_{L}$ and $t_R$ should be charged under this $U(1)_{E}$ and thus couple to the $Z'_{E}$ boson.\footnote{The $Z'_{E}$ can be even $t_R$-philic if the top Yukawa is generated through partial compositeness.} In this case, the evidence of the EHC mechanism can also be looked for through direct searches like $Z'_{E}\to t\bar{t}$ or $Z'_{E}\to b\bar{b}$. However, the dominant production comes from $b_L\bar{b}_L$ fusion with suppressed cross section and the two main channels both suffer from large backgrounds, which makes the constraint weak. Also, the heaviness of the top quark and the $b$-tag requirement make it hard to access the sub-TeV mass region, which is suggested by the Naturalness principle. So it turns out, for sub-TeV EHC bosons, the best way to look for them is through indirect searches, which will be discussed in Sec.~\ref{sec:Precision}.

\subsection{Cutoff of Higgs loop : Complex scalar singlet}

The cutoff of the Higgs loop, depending on its origin, is provided by the heavy gauge bosons and EHC gauge bosons discussed before. The dominant tree-level contribution is cut off by the complex scalar singlet $s$.

Its mass $M_s\sim \sqrt{c}\, g_{W'}f$ is expected to reside at few TeV. Its main interaction stems from the $s{H_2}^\dagger H_1$ coupling in Eq. \eqref{sHH} and the dominant visible decay channel is $hh$ with a branching ratio of $\sim 20\%$. There are also couplings with SM fermions due to mixing with the Higgs boson. Therefore, the production can be through similar channels as the Higgs production but is suppressed by the large $M_s$ and an additional factor of $\xi$ from mixing, which makes it hard to detect even at the HL-LHC. For example, taking $M_s=1300$ GeV, the $pp\to s \to hh$ cross section will be around 6$\xi$ fb $<0.6$ fb, far below the current constraint \cite{ATLAS:2022hwc,CMS:2021qvd} and also hard to access at the HL-LHC unless it is somewhat lighter \cite{ATLAS:2018crf,CidVidal:2018eel}.

\subsection{The Inert Second Higgs Doublet}

As discussed above, the second Higgs gets a similar potential as the first Higgs but without the top-loop contribution and its mass resides generically around a few hundred GeV. The gauge contribution will lead to a positive quadratic and quartic and thus no vev, $\langle H_{\rm inert}\rangle=0$, realizing a typical 'inert' Higgs doublet (see e.g.~\cite{Barbieri:2006dq,LopezHonorez:2006gr,Chowdhury:2011ga,Borah:2012pu,Cline:2013bln,Blinov:2015vma,Ilnicka:2015jba,Belyaev:2016lok,Fabian:2020hny,Astros:2023gda}).
Here, its components are expected to fulfill the relations
\begin{equation}
\begin{split}
M_{H} + M_{H^{\pm}}  > M_{W^{\pm}}\,,&\quad  
M_{A} + M_{H^{\pm}} > M_{W^{\pm}},\\
M_{H} + M_{A} > M_{Z}\,,&\quad 
2M_{H^{\pm}} > M_{Z}\,,
\end{split}
\end{equation}
as well as 
\begin{equation}
M_{H} > 80\,{\rm GeV},\ M_{H^\pm} > 70\,{\rm GeV}, 
\end{equation}
which assures that bounds from LEP searches for NP in EW gauge-boson decays and from LEP-II searches for direct production of the new scalars are easily evaded.

Moreover, similar to the vanilla inert doublet model, the neutral components of $H_{\rm inert}$ furnish attractive dark matter candidates. For a benchmark of $M_H \approx 500\,{\rm GeV}\lesssim M_{H^\pm} \lesssim M_A $, a realistic dark matter abundance could be achieved for small $\lambda_{g}+\frac{\lambda_s}{3}$, which assures that the annihilation into longitudinal vector bosons gets not too large, see, e.g., \cite{Belyaev:2016lok,Fabian:2020hny,Astros:2023gda}. We note that the small splitting to lift the $H^\pm$ components in order to avoid charged dark matter would need to originate from subleading contributions to the scalar potential. A detailed study of this is beyond the scope of this paper.

\subsection{Other pNGBs}

Finally, there is still the real singlet $\eta$, which is eaten by the $Z'_{H\!B}$ and results in a TeV-scale neutral boson. The coupling between $Z'_{H\!B}$ and the SM fermions is determined by their charge and depends on the concrete extension of the $U(1)_{H\!B}$ symmetry. One possible setup is discussed in ref.
\cite{Chung:2021ekz,Chung:2021xhd}, where the $Z'_{H\!B}$ is third generation philic and the production rate is highly suppressed.



\section{Precision Tests}\label{sec:Precision}

In this section, we discuss the indirect tests of the model from precision measurements, 
focusing on the
\mbox{dim-6} top Yukawa coupling and the heavy~$SU(2)'$~bosons. 

\subsection{Running Top Yukawa}\label{sec:ytpheno}

The top Yukawa coupling from the four fermion interaction is one of the main characteristic features in our model. The low cutoff implies large running at the corresponding mass scale, which will reveal itself in differential momentum
distributions of $t\bar{t}h$ production \cite{MammenAbraham:2021ssc,Bittar:2022wgb}. However, based on the $t\bar{t}h$ cross section, it is hard to achieve the desired sensitivity in the LHC era and there are currently no analyses available.

\subsection{Running Top mass}\label{sec:mtpheno}

Besides probing the top Yukawa at high scales directly, the model can be tested by measuring the running of the top quark mass, which will affect the $t\bar{t}$ differential cross section. Without a Higgs in the final state, the cross section is much larger. The measurement has already been performed by CMS using 35.9 fb$^{-1}$ of run 2 data~\cite{CMS:2019jul} and has been interpreted in \cite{Defranchis:2022nqb} in terms of the top mass running up to $0.5$ TeV. This already puts a bound of $M_{E}\gtrsim700$ GeV \cite{Chung:2023iwj}, still in good agreement with a natural model, see Sec.\ref{sec:VH}. With more data coming out, we however expect that the fine-tuning region down to $10\%$ can be tested in the HL-LHC era.

\subsection{Four top quarks cross section}\label{sec:4tpheno}

The $Z'_{E}$ boson, even though hard to search for if its mass is below $1$ TeV, will have a direct impact on the four top-quark cross section. In the SM, the prediction for the cross section was updated with next-to-leading logarithmic (NLL') accuracy \cite{vanBeekveld:2022hty}. From the experimental side, both ATLAS \cite{ATLAS:2023ajo} and CMS \cite{CMS:2023ftu} have just reached the $5\,\sigma$ observation level with LHC run 2 data, which also allows us to put a constraint on the scenario. Several analyses aiming at interpreting the results in terms of simplified models or effective theories have been performed in recent years \cite{Darme:2021gtt, Banelli:2020iau, Blekman:2022jag}. Following the simplified-model analysis of~\cite{Darme:2021gtt}, we obtain mild constraints of $M_{Z'_{E}}\gtrsim 0.5$\,TeV for $g_{Z'_{E}}=0.3$ and $M_{Z'_{E}}\gtrsim 1$\,TeV for $g_{Z'_{E}}=1.0$ at~95\%~C.L.\,.\footnote{Notice that the $Z'_{E}$ coupling to top quark can be different from $g_{E}$ up to a factor depending on the charge. The mass $M_{Z'_{E}}$ can also be different from $M_E$ of the $E_\mu$ bosons up to an $O(1)$ factor.}

\subsection{Flavor constraints}\label{sec:Flavor}

The $Z'_{E}$ boson can also have an impact on light-quark physics through mixing, which can introduce dangerous flavor changing neutral currents. Assuming $\theta_{sb}\gg\theta_{db}$ for the down-quark mixing, in analogy to the CKM matrix, the strongest constraint is expected from $B_s-\bar{B}_s$ mixing, which contains second and third generation quarks. Following \cite{DiLuzio:2017fdq}
we arive at the 95\% C.L. bound~\cite{Allanach:2019mfl}
\begin{equation}\label{Bsmixing}
\frac{g_{sb}}{M_{Z'}}=\frac{g_{Z^\prime_E}}{M_{Z^\prime_E}} \theta_{sb} \leq \frac{1}{194~\text{TeV}}\,,
\end{equation}
where the angle $\theta_{sb}$ parametrizes the rotation between the second and third generations of down-type quarks in the mass basis and ${g_{Z^\prime_E}}/{M_{Z^\prime_E}}\approx 1/f_{E}$. For $f_{E}=4$ TeV we arrive at $\theta_{sb} \lesssim 0.02$.
Applying the same constraint to the $Z'$ of $SU(2)'$, where ${g_{sb}}/{M_{Z'}}={g_{W'}}\theta_{sb}/2{M_{W'}}=\theta_{sb}/f$, induces a stronger bound of $\theta_{sb}\lesssim0.01$.


\subsection{Electroweak precision tests}\label{sec:EWPT}

Adding new degrees of freedom at the TeV scale will lead to electroweak oblique corrections, which are usually expressed in terms of $S$, $T$, and $U$ parameters~\cite{Peskin:1990zt,Peskin:1991sw}. The current global fit results in (fixing $U=0$)~\cite{ParticleDataGroup:2022pth}
\begin{equation}
S=-0.01 \pm 0.07, \quad T=0.04 \pm 0.06,
\end{equation}
with a strong positive correlation (92\%) between them. At 95\% C.L., one obtains $S<0.14$ and $T<0.22$.

There are several contributions to the oblique parameters in our model. In CHMs, the nonlinear Higgs dynamics will contribute to both $S$ and $T$ due to the modifications of the Higgs couplings~\cite{Barbieri:2007bh,Grojean:2013qca}, which is proportional to $\xi$ and depends logarithmically on $M_\rho/m_h$ as 
\begin{align}
\Delta S=\frac{1}{12\pi}\text{ ln}\left(\frac{M_\rho^2}{m_h^2}\right)\cdot\xi\,,~
\Delta T=-\frac{3}{16\pi c_w^2}\text{ ln}\left(\frac{M_\rho^2}{m_h^2}\right)\cdot\xi\,.
\end{align}
Taking $M_\rho=5$~TeV, we get
\begin{align}\label{ST1}
\Delta S\sim 0.20 ~\xi\,,\quad \Delta T\sim -0.57 ~\xi~.
\end{align}
The other contributions are discussed below.

\subsubsection*{The $S$ parameter} 

In FCHMs, the hyperfermions, which carry electroweak quantum numbers, also contribute to the $S$ parameter. The contribution can be estimated by calculating the one-loop diagram from the heavy  hyperfermions. In this case, each heavy $SU(2)_W$ doublet adds $\sim \xi/6\pi$, resulting in
\begin{align}\label{ST2}
\Delta S\sim \frac{\xi}{6\pi}\,2N_{HC}\sim 0.21~\xi \quad(N_{HC}=2)~.
\end{align}

In little Higgs models, the $S$ parameter also receives a contribution from the heavy $SU(2)'$ bosons, which dominates in our setup, reading \cite{Csaki:2002qg, Hewett:2002px,Csaki:2003si,Gregoire:2003kr,Marandella:2005wd,Strumia:2022qkt}
\begin{align}
\Delta S \sim \frac{8\pi v^2}{g_{W}^2f^2}\text{ sin}^2\,{ \theta_e} \sim  \frac{8\pi}{g_{W'}^2}\,\xi = 2\pi \frac{v^2}{M_{W'}^2}\,.
\end{align}
This leads to a bound of $M_{W'}\gtrsim 1.6$\,TeV, being comparable with collider constraints discussed above.

\subsubsection*{The $T$ parameter} 

The $SU(2)'$ bosons contribute to the $T$ parameter only at $O(\xi^2)$, leading to a small negative correction of~\cite{Chiang:2009kb}
\begin{align}
\Delta T=-\frac{\text{sin}^2\,{ \theta_W}}{4 \alpha \,\text{cos}^2\,{ \theta_W}}\,\text{sin}^2\,{ \theta_e}\,\text{cos}^2\,{ \theta_e}\frac{v^4}{f^4}~.
\end{align}
Also, a generic 2HDM of $SU(6)/Sp(6)$ induces~\cite{Csaki:2003si,Gregoire:2003kr,Marandella:2005wd,Strumia:2022qkt}
\begin{align}
\Delta T= \frac{1}{\alpha}\frac{v^2}{4f^2} \text{ cos}^2 \,2\beta~,
\end{align}
which vanishes in our model because tan$\,\beta=1$ preserves custodial symmetry. Yet another contribution comes from the hyperfermion loop with an insertion of an EHC boson, treating top and bottom quarks differently,
\begin{align}
\Delta T= \frac{1}{\alpha}\frac{N_c Y_s^4}{(16\pi^2)^2}\frac{v^2}{f_E^2}~,
\end{align}
which provides a small positive contribution leading in total to a negligible $\Delta T$.

\subsection{$Zb\bar{b}$ coupling}\label{sec:Zbb}

Another relevant constraint on top quark related models comes from the $Zb\bar{b}$ coupling. The deviation for the left-handed bottom coupling is constrained within $-2.5\times 10^{-3} \lesssim \delta g_{b_L} \lesssim 5\times 10^{-3}$~\cite{Batell:2012ca, Guadagnoli:2013mru} at $95\%$ C.L., while $0 \lesssim \delta g_{b_R} \lesssim 3\times 10^{-2}$ is much less constrained.

Here, the correction from mixing with~$SU(2)'$~bosons, 
\begin{align}
\delta g_{b_L} = -\frac{1}{4}\,\text{cos}^4\,{\theta_e}\,\text{sin}^2\beta\,\frac{v^2}{f^2}~,
\end{align}
is partially cancelled by the hyperfermion's effect~\cite{Chivukula:1992ap}
\begin{align}
\delta g_{b_L} = \frac{1}{4}\frac{v^2}{f_{E}^2}\sim \frac{1}{4}\frac{y_t}{Y_s}\frac{v^2}{f^2}~,
\end{align}
which in total leads to a correction well within bounds.

\section{Conclusions} \label{sec:Conclusion}

We presented a model that cuts off the dangerous corrections to the Higgs mass term via third-generation-philic bosons that can be light while escaping strong experimental limits. Interestingly, the corresponding states, a $W^\prime$, a light EHC boson, and a scalar singlet, can naturally emerge in straightforward incarnations of a (composite) Goldstone-Higgs.

As demonstrated, experiments are just starting to explore the natural parameter region of the model. On the other hand, at the HL-LHC we expect that tuning of down to 10\% will be tested, while only future colliders such as FCC can exclude this scenario of hidden Naturalness to a point of excessive tuning well below 10\%.

\section*{Acknowledgments}

We thank Andreas Bally, Hsin-Chia Cheng and Matthias Neubert for useful discussions and comments.

\newpage

\bibliography{3G_Ref}

\begin{thebibliography}{81}%
\makeatletter
\providecommand \@ifxundefined [1]{%
 \@ifx{#1\undefined}
}%
\providecommand \@ifnum [1]{%
 \ifnum #1\expandafter \@firstoftwo
 \else \expandafter \@secondoftwo
 \fi
}%
\providecommand \@ifx [1]{%
 \ifx #1\expandafter \@firstoftwo
 \else \expandafter \@secondoftwo
 \fi
}%
\providecommand \natexlab [1]{#1}%
\providecommand \enquote  [1]{``#1''}%
\providecommand \bibnamefont  [1]{#1}%
\providecommand \bibfnamefont [1]{#1}%
\providecommand \citenamefont [1]{#1}%
\providecommand \href@noop [0]{\@secondoftwo}%
\providecommand \href [0]{\begingroup \@sanitize@url \@href}%
\providecommand \@href[1]{\@@startlink{#1}\@@href}%
\providecommand \@@href[1]{\endgroup#1\@@endlink}%
\providecommand \@sanitize@url [0]{\catcode `\\12\catcode `\$12\catcode
  `\&12\catcode `\#12\catcode `\^12\catcode `\_12\catcode `\%12\relax}%
\providecommand \@@startlink[1]{}%
\providecommand \@@endlink[0]{}%
\providecommand \url  [0]{\begingroup\@sanitize@url \@url }%
\providecommand \@url [1]{\endgroup\@href {#1}{\urlprefix }}%
\providecommand \urlprefix  [0]{URL }%
\providecommand \Eprint [0]{\href }%
\providecommand \doibase [0]{http://dx.doi.org/}%
\providecommand \selectlanguage [0]{\@gobble}%
\providecommand \bibinfo  [0]{\@secondoftwo}%
\providecommand \bibfield  [0]{\@secondoftwo}%
\providecommand \translation [1]{[#1]}%
\providecommand \BibitemOpen [0]{}%
\providecommand \bibitemStop [0]{}%
\providecommand \bibitemNoStop [0]{.\EOS\space}%
\providecommand \EOS [0]{\spacefactor3000\relax}%
\providecommand \BibitemShut  [1]{\csname bibitem#1\endcsname}%
\let\auto@bib@innerbib\@empty
\bibitem [{\citenamefont {Chatrchyan}\ \emph {et~al.}(2012)\citenamefont
  {Chatrchyan} \emph {et~al.}}]{Chatrchyan:2012xdj}%
  \BibitemOpen
  \bibfield  {author} {\bibinfo {author} {\bibfnamefont {S.}~\bibnamefont
  {Chatrchyan}} \emph {et~al.} (\bibinfo {collaboration} {CMS}),\ }\href
  {\doibase 10.1016/j.physletb.2012.08.021} {\bibfield  {journal} {\bibinfo
  {journal} {Phys. Lett.}\ }\textbf {\bibinfo {volume} {B716}},\ \bibinfo
  {pages} {30} (\bibinfo {year} {2012})},\ \Eprint
  {http://arxiv.org/abs/1207.7235} {arXiv:1207.7235 [hep-ex]} \BibitemShut
  {NoStop}%
\bibitem [{\citenamefont {Aad}\ \emph {et~al.}(2012)\citenamefont {Aad} \emph
  {et~al.}}]{Aad:2012tfa}%
  \BibitemOpen
  \bibfield  {author} {\bibinfo {author} {\bibfnamefont {G.}~\bibnamefont
  {Aad}} \emph {et~al.} (\bibinfo {collaboration} {ATLAS}),\ }\href {\doibase
  10.1016/j.physletb.2012.08.020} {\bibfield  {journal} {\bibinfo  {journal}
  {Phys. Lett.}\ }\textbf {\bibinfo {volume} {B716}},\ \bibinfo {pages} {1}
  (\bibinfo {year} {2012})},\ \Eprint {http://arxiv.org/abs/1207.7214}
  {arXiv:1207.7214 [hep-ex]} \BibitemShut {NoStop}%
\bibitem [{\citenamefont {Chacko}\ \emph {et~al.}(2006)\citenamefont {Chacko},
  \citenamefont {Goh},\ and\ \citenamefont {Harnik}}]{Chacko:2005pe}%
  \BibitemOpen
  \bibfield  {author} {\bibinfo {author} {\bibfnamefont {Z.}~\bibnamefont
  {Chacko}}, \bibinfo {author} {\bibfnamefont {H.-S.}\ \bibnamefont {Goh}}, \
  and\ \bibinfo {author} {\bibfnamefont {R.}~\bibnamefont {Harnik}},\ }\href
  {\doibase 10.1103/PhysRevLett.96.231802} {\bibfield  {journal} {\bibinfo
  {journal} {Phys. Rev. Lett.}\ }\textbf {\bibinfo {volume} {96}},\ \bibinfo
  {pages} {231802} (\bibinfo {year} {2006})},\ \Eprint
  {http://arxiv.org/abs/hep-ph/0506256} {arXiv:hep-ph/0506256 [hep-ph]}
  \BibitemShut {NoStop}%
\bibitem [{\citenamefont {Barbieri}\ \emph {et~al.}(2011)\citenamefont
  {Barbieri}, \citenamefont {Isidori}, \citenamefont {Jones-Perez},
  \citenamefont {Lodone},\ and\ \citenamefont {Straub}}]{Barbieri:2011ci}%
  \BibitemOpen
  \bibfield  {author} {\bibinfo {author} {\bibfnamefont {R.}~\bibnamefont
  {Barbieri}}, \bibinfo {author} {\bibfnamefont {G.}~\bibnamefont {Isidori}},
  \bibinfo {author} {\bibfnamefont {J.}~\bibnamefont {Jones-Perez}}, \bibinfo
  {author} {\bibfnamefont {P.}~\bibnamefont {Lodone}}, \ and\ \bibinfo {author}
  {\bibfnamefont {D.~M.}\ \bibnamefont {Straub}},\ }\href {\doibase
  10.1140/epjc/s10052-011-1725-z} {\bibfield  {journal} {\bibinfo  {journal}
  {Eur. Phys. J. C}\ }\textbf {\bibinfo {volume} {71}},\ \bibinfo {pages}
  {1725} (\bibinfo {year} {2011})},\ \Eprint {http://arxiv.org/abs/1105.2296}
  {arXiv:1105.2296 [hep-ph]} \BibitemShut {NoStop}%
\bibitem [{\citenamefont {Isidori}\ and\ \citenamefont
  {Straub}(2012)}]{Isidori:2012ts}%
  \BibitemOpen
  \bibfield  {author} {\bibinfo {author} {\bibfnamefont {G.}~\bibnamefont
  {Isidori}}\ and\ \bibinfo {author} {\bibfnamefont {D.~M.}\ \bibnamefont
  {Straub}},\ }\href {\doibase 10.1140/epjc/s10052-012-2103-1} {\bibfield
  {journal} {\bibinfo  {journal} {Eur. Phys. J. C}\ }\textbf {\bibinfo {volume}
  {72}},\ \bibinfo {pages} {2103} (\bibinfo {year} {2012})},\ \Eprint
  {http://arxiv.org/abs/1202.0464} {arXiv:1202.0464 [hep-ph]} \BibitemShut
  {NoStop}%
\bibitem [{\citenamefont {Cacciapaglia}\ and\ \citenamefont
  {Sannino}(2014)}]{Cacciapaglia:2014uja}%
  \BibitemOpen
  \bibfield  {author} {\bibinfo {author} {\bibfnamefont {G.}~\bibnamefont
  {Cacciapaglia}}\ and\ \bibinfo {author} {\bibfnamefont {F.}~\bibnamefont
  {Sannino}},\ }\href {\doibase 10.1007/JHEP04(2014)111} {\bibfield  {journal}
  {\bibinfo  {journal} {JHEP}\ }\textbf {\bibinfo {volume} {04}},\ \bibinfo
  {pages} {111} (\bibinfo {year} {2014})},\ \Eprint
  {http://arxiv.org/abs/1402.0233} {arXiv:1402.0233 [hep-ph]} \BibitemShut
  {NoStop}%
\bibitem [{\citenamefont {Cacciapaglia}\ \emph {et~al.}(2020)\citenamefont
  {Cacciapaglia}, \citenamefont {Pica},\ and\ \citenamefont
  {Sannino}}]{Cacciapaglia:2020kgq}%
  \BibitemOpen
  \bibfield  {author} {\bibinfo {author} {\bibfnamefont {G.}~\bibnamefont
  {Cacciapaglia}}, \bibinfo {author} {\bibfnamefont {C.}~\bibnamefont {Pica}},
  \ and\ \bibinfo {author} {\bibfnamefont {F.}~\bibnamefont {Sannino}},\ }\href
  {\doibase 10.1016/j.physrep.2020.07.002} {\bibfield  {journal} {\bibinfo
  {journal} {Phys. Rept.}\ }\textbf {\bibinfo {volume} {877}},\ \bibinfo
  {pages} {1} (\bibinfo {year} {2020})},\ \Eprint
  {http://arxiv.org/abs/2002.04914} {arXiv:2002.04914 [hep-ph]} \BibitemShut
  {NoStop}%
\bibitem [{\citenamefont {Arkani-Hamed}\ \emph {et~al.}(2001)\citenamefont
  {Arkani-Hamed}, \citenamefont {Cohen},\ and\ \citenamefont
  {Georgi}}]{ArkaniHamed:2001nc}%
  \BibitemOpen
  \bibfield  {author} {\bibinfo {author} {\bibfnamefont {N.}~\bibnamefont
  {Arkani-Hamed}}, \bibinfo {author} {\bibfnamefont {A.~G.}\ \bibnamefont
  {Cohen}}, \ and\ \bibinfo {author} {\bibfnamefont {H.}~\bibnamefont
  {Georgi}},\ }\href {\doibase 10.1016/S0370-2693(01)00741-9} {\bibfield
  {journal} {\bibinfo  {journal} {Phys. Lett.}\ }\textbf {\bibinfo {volume}
  {B513}},\ \bibinfo {pages} {232} (\bibinfo {year} {2001})},\ \Eprint
  {http://arxiv.org/abs/hep-ph/0105239} {arXiv:hep-ph/0105239 [hep-ph]}
  \BibitemShut {NoStop}%
\bibitem [{\citenamefont {Arkani-Hamed}\ \emph
  {et~al.}(2002{\natexlab{a}})\citenamefont {Arkani-Hamed}, \citenamefont
  {Cohen}, \citenamefont {Katz}, \citenamefont {Nelson}, \citenamefont
  {Gregoire},\ and\ \citenamefont {Wacker}}]{ArkaniHamed:2002qx}%
  \BibitemOpen
  \bibfield  {author} {\bibinfo {author} {\bibfnamefont {N.}~\bibnamefont
  {Arkani-Hamed}}, \bibinfo {author} {\bibfnamefont {A.~G.}\ \bibnamefont
  {Cohen}}, \bibinfo {author} {\bibfnamefont {E.}~\bibnamefont {Katz}},
  \bibinfo {author} {\bibfnamefont {A.~E.}\ \bibnamefont {Nelson}}, \bibinfo
  {author} {\bibfnamefont {T.}~\bibnamefont {Gregoire}}, \ and\ \bibinfo
  {author} {\bibfnamefont {J.~G.}\ \bibnamefont {Wacker}},\ }\href {\doibase
  10.1088/1126-6708/2002/08/021} {\bibfield  {journal} {\bibinfo  {journal}
  {JHEP}\ }\textbf {\bibinfo {volume} {08}},\ \bibinfo {pages} {021} (\bibinfo
  {year} {2002}{\natexlab{a}})},\ \Eprint {http://arxiv.org/abs/hep-ph/0206020}
  {arXiv:hep-ph/0206020 [hep-ph]} \BibitemShut {NoStop}%
\bibitem [{\citenamefont {Arkani-Hamed}\ \emph
  {et~al.}(2002{\natexlab{b}})\citenamefont {Arkani-Hamed}, \citenamefont
  {Cohen}, \citenamefont {Katz},\ and\ \citenamefont
  {Nelson}}]{ArkaniHamed:2002qy}%
  \BibitemOpen
  \bibfield  {author} {\bibinfo {author} {\bibfnamefont {N.}~\bibnamefont
  {Arkani-Hamed}}, \bibinfo {author} {\bibfnamefont {A.~G.}\ \bibnamefont
  {Cohen}}, \bibinfo {author} {\bibfnamefont {E.}~\bibnamefont {Katz}}, \ and\
  \bibinfo {author} {\bibfnamefont {A.~E.}\ \bibnamefont {Nelson}},\ }\href
  {\doibase 10.1088/1126-6708/2002/07/034} {\bibfield  {journal} {\bibinfo
  {journal} {JHEP}\ }\textbf {\bibinfo {volume} {07}},\ \bibinfo {pages} {034}
  (\bibinfo {year} {2002}{\natexlab{b}})},\ \Eprint
  {http://arxiv.org/abs/hep-ph/0206021} {arXiv:hep-ph/0206021 [hep-ph]}
  \BibitemShut {NoStop}%
\bibitem [{\citenamefont {Low}\ \emph {et~al.}(2002)\citenamefont {Low},
  \citenamefont {Skiba},\ and\ \citenamefont {Tucker-Smith}}]{Low:2002ws}%
  \BibitemOpen
  \bibfield  {author} {\bibinfo {author} {\bibfnamefont {I.}~\bibnamefont
  {Low}}, \bibinfo {author} {\bibfnamefont {W.}~\bibnamefont {Skiba}}, \ and\
  \bibinfo {author} {\bibfnamefont {D.}~\bibnamefont {Tucker-Smith}},\ }\href
  {\doibase 10.1103/PhysRevD.66.072001} {\bibfield  {journal} {\bibinfo
  {journal} {Phys. Rev.}\ }\textbf {\bibinfo {volume} {D66}},\ \bibinfo {pages}
  {072001} (\bibinfo {year} {2002})},\ \Eprint
  {http://arxiv.org/abs/hep-ph/0207243} {arXiv:hep-ph/0207243 [hep-ph]}
  \BibitemShut {NoStop}%
\bibitem [{\citenamefont {Barnard}\ \emph {et~al.}(2014)\citenamefont
  {Barnard}, \citenamefont {Gherghetta},\ and\ \citenamefont
  {Ray}}]{Barnard:2013zea}%
  \BibitemOpen
  \bibfield  {author} {\bibinfo {author} {\bibfnamefont {J.}~\bibnamefont
  {Barnard}}, \bibinfo {author} {\bibfnamefont {T.}~\bibnamefont {Gherghetta}},
  \ and\ \bibinfo {author} {\bibfnamefont {T.~S.}\ \bibnamefont {Ray}},\ }\href
  {\doibase 10.1007/JHEP02(2014)002} {\bibfield  {journal} {\bibinfo  {journal}
  {JHEP}\ }\textbf {\bibinfo {volume} {02}},\ \bibinfo {pages} {002} (\bibinfo
  {year} {2014})},\ \Eprint {http://arxiv.org/abs/1311.6562} {arXiv:1311.6562
  [hep-ph]} \BibitemShut {NoStop}%
\bibitem [{\citenamefont {Ferretti}\ and\ \citenamefont
  {Karateev}(2014)}]{Ferretti:2013kya}%
  \BibitemOpen
  \bibfield  {author} {\bibinfo {author} {\bibfnamefont {G.}~\bibnamefont
  {Ferretti}}\ and\ \bibinfo {author} {\bibfnamefont {D.}~\bibnamefont
  {Karateev}},\ }\href {\doibase 10.1007/JHEP03(2014)077} {\bibfield  {journal}
  {\bibinfo  {journal} {JHEP}\ }\textbf {\bibinfo {volume} {03}},\ \bibinfo
  {pages} {077} (\bibinfo {year} {2014})},\ \Eprint
  {http://arxiv.org/abs/1312.5330} {arXiv:1312.5330 [hep-ph]} \BibitemShut
  {NoStop}%
\bibitem [{\citenamefont {Cacciapaglia}\ and\ \citenamefont
  {Sannino}(2016)}]{Cacciapaglia:2015yra}%
  \BibitemOpen
  \bibfield  {author} {\bibinfo {author} {\bibfnamefont {G.}~\bibnamefont
  {Cacciapaglia}}\ and\ \bibinfo {author} {\bibfnamefont {F.}~\bibnamefont
  {Sannino}},\ }\href {\doibase 10.1016/j.physletb.2016.02.034} {\bibfield
  {journal} {\bibinfo  {journal} {Phys. Lett. B}\ }\textbf {\bibinfo {volume}
  {755}},\ \bibinfo {pages} {328} (\bibinfo {year} {2016})},\ \Eprint
  {http://arxiv.org/abs/1508.00016} {arXiv:1508.00016 [hep-ph]} \BibitemShut
  {NoStop}%
\bibitem [{\citenamefont {Bally}\ \emph
  {et~al.}(2023{\natexlab{a}})\citenamefont {Bally}, \citenamefont {Chung},\
  and\ \citenamefont {Goertz}}]{Bally:2023lji}%
  \BibitemOpen
  \bibfield  {author} {\bibinfo {author} {\bibfnamefont {A.}~\bibnamefont
  {Bally}}, \bibinfo {author} {\bibfnamefont {Y.}~\bibnamefont {Chung}}, \ and\
  \bibinfo {author} {\bibfnamefont {F.}~\bibnamefont {Goertz}},\ }in\
  \href@noop {} {\emph {\bibinfo {booktitle} {{57th Rencontres de Moriond on
  QCD and High Energy Interactions}}}}\ (\bibinfo {year} {2023})\ \Eprint
  {http://arxiv.org/abs/2304.11891} {arXiv:2304.11891 [hep-ph]} \BibitemShut
  {NoStop}%
\bibitem [{\citenamefont {Chung}(2023)}]{Chung:2023iwj}%
  \BibitemOpen
  \bibfield  {author} {\bibinfo {author} {\bibfnamefont {Y.}~\bibnamefont
  {Chung}},\ }\href@noop {} {\  (\bibinfo {year} {2023})},\ \Eprint
  {http://arxiv.org/abs/2309.00072} {arXiv:2309.00072 [hep-ph]} \BibitemShut
  {NoStop}%
\bibitem [{\citenamefont {Bally}\ \emph
  {et~al.}(2023{\natexlab{b}})\citenamefont {Bally}, \citenamefont {Chung},\
  and\ \citenamefont {Goertz}}]{Bally:2022naz}%
  \BibitemOpen
  \bibfield  {author} {\bibinfo {author} {\bibfnamefont {A.}~\bibnamefont
  {Bally}}, \bibinfo {author} {\bibfnamefont {Y.}~\bibnamefont {Chung}}, \ and\
  \bibinfo {author} {\bibfnamefont {F.}~\bibnamefont {Goertz}},\ }\href
  {\doibase 10.1103/PhysRevD.108.055008} {\bibfield  {journal} {\bibinfo
  {journal} {Phys. Rev. D}\ }\textbf {\bibinfo {volume} {108}},\ \bibinfo
  {pages} {055008} (\bibinfo {year} {2023}{\natexlab{b}})},\ \Eprint
  {http://arxiv.org/abs/2211.17254} {arXiv:2211.17254 [hep-ph]} \BibitemShut
  {NoStop}%
\bibitem [{\citenamefont {Malkawi}\ \emph {et~al.}(1996)\citenamefont
  {Malkawi}, \citenamefont {Tait},\ and\ \citenamefont
  {Yuan}}]{Malkawi:1996fs}%
  \BibitemOpen
  \bibfield  {author} {\bibinfo {author} {\bibfnamefont {E.}~\bibnamefont
  {Malkawi}}, \bibinfo {author} {\bibfnamefont {T.~M.~P.}\ \bibnamefont
  {Tait}}, \ and\ \bibinfo {author} {\bibfnamefont {C.~P.}\ \bibnamefont
  {Yuan}},\ }\href {\doibase 10.1016/0370-2693(96)00859-3} {\bibfield
  {journal} {\bibinfo  {journal} {Phys. Lett. B}\ }\textbf {\bibinfo {volume}
  {385}},\ \bibinfo {pages} {304} (\bibinfo {year} {1996})},\ \Eprint
  {http://arxiv.org/abs/hep-ph/9603349} {arXiv:hep-ph/9603349} \BibitemShut
  {NoStop}%
\bibitem [{\citenamefont {Muller}\ and\ \citenamefont
  {Nandi}(1996)}]{Muller:1996dj}%
  \BibitemOpen
  \bibfield  {author} {\bibinfo {author} {\bibfnamefont {D.~J.}\ \bibnamefont
  {Muller}}\ and\ \bibinfo {author} {\bibfnamefont {S.}~\bibnamefont {Nandi}},\
  }\href {\doibase 10.1016/0370-2693(96)00745-9} {\bibfield  {journal}
  {\bibinfo  {journal} {Phys. Lett. B}\ }\textbf {\bibinfo {volume} {383}},\
  \bibinfo {pages} {345} (\bibinfo {year} {1996})},\ \Eprint
  {http://arxiv.org/abs/hep-ph/9602390} {arXiv:hep-ph/9602390} \BibitemShut
  {NoStop}%
\bibitem [{\citenamefont {Chiang}\ \emph {et~al.}(2010)\citenamefont {Chiang},
  \citenamefont {Deshpande}, \citenamefont {He},\ and\ \citenamefont
  {Jiang}}]{Chiang:2009kb}%
  \BibitemOpen
  \bibfield  {author} {\bibinfo {author} {\bibfnamefont {C.-W.}\ \bibnamefont
  {Chiang}}, \bibinfo {author} {\bibfnamefont {N.~G.}\ \bibnamefont
  {Deshpande}}, \bibinfo {author} {\bibfnamefont {X.-G.}\ \bibnamefont {He}}, \
  and\ \bibinfo {author} {\bibfnamefont {J.}~\bibnamefont {Jiang}},\ }\href
  {\doibase 10.1103/PhysRevD.81.015006} {\bibfield  {journal} {\bibinfo
  {journal} {Phys. Rev. D}\ }\textbf {\bibinfo {volume} {81}},\ \bibinfo
  {pages} {015006} (\bibinfo {year} {2010})},\ \Eprint
  {http://arxiv.org/abs/0911.1480} {arXiv:0911.1480 [hep-ph]} \BibitemShut
  {NoStop}%
\bibitem [{\citenamefont {Katz}\ \emph {et~al.}(2005)\citenamefont {Katz},
  \citenamefont {Nelson},\ and\ \citenamefont {Walker}}]{Katz:2005au}%
  \BibitemOpen
  \bibfield  {author} {\bibinfo {author} {\bibfnamefont {E.}~\bibnamefont
  {Katz}}, \bibinfo {author} {\bibfnamefont {A.~E.}\ \bibnamefont {Nelson}}, \
  and\ \bibinfo {author} {\bibfnamefont {D.~G.~E.}\ \bibnamefont {Walker}},\
  }\href {\doibase 10.1088/1126-6708/2005/08/074} {\bibfield  {journal}
  {\bibinfo  {journal} {JHEP}\ }\textbf {\bibinfo {volume} {08}},\ \bibinfo
  {pages} {074} (\bibinfo {year} {2005})},\ \Eprint
  {http://arxiv.org/abs/hep-ph/0504252} {arXiv:hep-ph/0504252} \BibitemShut
  {NoStop}%
\bibitem [{\citenamefont {Cai}\ \emph {et~al.}(2020)\citenamefont {Cai},
  \citenamefont {Zhang}, \citenamefont {Cacciapaglia}, \citenamefont
  {Rosenlyst},\ and\ \citenamefont {Frandsen}}]{Cai:2019cow}%
  \BibitemOpen
  \bibfield  {author} {\bibinfo {author} {\bibfnamefont {C.}~\bibnamefont
  {Cai}}, \bibinfo {author} {\bibfnamefont {H.-H.}\ \bibnamefont {Zhang}},
  \bibinfo {author} {\bibfnamefont {G.}~\bibnamefont {Cacciapaglia}}, \bibinfo
  {author} {\bibfnamefont {M.}~\bibnamefont {Rosenlyst}}, \ and\ \bibinfo
  {author} {\bibfnamefont {M.~T.}\ \bibnamefont {Frandsen}},\ }\href {\doibase
  10.1103/PhysRevLett.125.021801} {\bibfield  {journal} {\bibinfo  {journal}
  {Phys. Rev. Lett.}\ }\textbf {\bibinfo {volume} {125}},\ \bibinfo {pages}
  {021801} (\bibinfo {year} {2020})},\ \Eprint
  {http://arxiv.org/abs/1911.12130} {arXiv:1911.12130 [hep-ph]} \BibitemShut
  {NoStop}%
\bibitem [{\citenamefont {Rosenlyst}\ and\ \citenamefont
  {Hill}(2020)}]{Rosenlyst:2020znn}%
  \BibitemOpen
  \bibfield  {author} {\bibinfo {author} {\bibfnamefont {M.}~\bibnamefont
  {Rosenlyst}}\ and\ \bibinfo {author} {\bibfnamefont {C.~T.}\ \bibnamefont
  {Hill}},\ }\href {\doibase 10.1103/PhysRevD.101.095027} {\bibfield  {journal}
  {\bibinfo  {journal} {Phys. Rev. D}\ }\textbf {\bibinfo {volume} {101}},\
  \bibinfo {pages} {095027} (\bibinfo {year} {2020})},\ \Eprint
  {http://arxiv.org/abs/2002.04931} {arXiv:2002.04931 [hep-ph]} \BibitemShut
  {NoStop}%
\bibitem [{\citenamefont {Cheng}\ and\ \citenamefont
  {Chung}(2020)}]{Cheng:2020dum}%
  \BibitemOpen
  \bibfield  {author} {\bibinfo {author} {\bibfnamefont {H.-C.}\ \bibnamefont
  {Cheng}}\ and\ \bibinfo {author} {\bibfnamefont {Y.}~\bibnamefont {Chung}},\
  }\href {\doibase 10.1007/JHEP10(2020)175} {\bibfield  {journal} {\bibinfo
  {journal} {JHEP}\ }\textbf {\bibinfo {volume} {10}},\ \bibinfo {pages} {175}
  (\bibinfo {year} {2020})},\ \Eprint {http://arxiv.org/abs/2007.11780}
  {arXiv:2007.11780 [hep-ph]} \BibitemShut {NoStop}%
\bibitem [{\citenamefont {Cacciapaglia}\ and\ \citenamefont
  {Rosenlyst}(2021)}]{Cacciapaglia:2020psm}%
  \BibitemOpen
  \bibfield  {author} {\bibinfo {author} {\bibfnamefont {G.}~\bibnamefont
  {Cacciapaglia}}\ and\ \bibinfo {author} {\bibfnamefont {M.}~\bibnamefont
  {Rosenlyst}},\ }\href {\doibase 10.1007/JHEP09(2021)167} {\bibfield
  {journal} {\bibinfo  {journal} {JHEP}\ }\textbf {\bibinfo {volume} {09}},\
  \bibinfo {pages} {167} (\bibinfo {year} {2021})},\ \Eprint
  {http://arxiv.org/abs/2010.01437} {arXiv:2010.01437 [hep-ph]} \BibitemShut
  {NoStop}%
\bibitem [{\citenamefont {Chung}(2021{\natexlab{a}})}]{Chung:2021fpc}%
  \BibitemOpen
  \bibfield  {author} {\bibinfo {author} {\bibfnamefont {Y.}~\bibnamefont
  {Chung}},\ }\href {\doibase 10.1103/PhysRevD.104.095011} {\bibfield
  {journal} {\bibinfo  {journal} {Phys. Rev. D}\ }\textbf {\bibinfo {volume}
  {104}},\ \bibinfo {pages} {095011} (\bibinfo {year} {2021}{\natexlab{a}})},\
  \Eprint {http://arxiv.org/abs/2104.11719} {arXiv:2104.11719 [hep-ph]}
  \BibitemShut {NoStop}%
\bibitem [{\citenamefont {Rosenlyst}(2022)}]{Rosenlyst:2021tdr}%
  \BibitemOpen
  \bibfield  {author} {\bibinfo {author} {\bibfnamefont {M.}~\bibnamefont
  {Rosenlyst}},\ }\href {\doibase 10.1103/PhysRevD.106.013002} {\bibfield
  {journal} {\bibinfo  {journal} {Phys. Rev. D}\ }\textbf {\bibinfo {volume}
  {106}},\ \bibinfo {pages} {013002} (\bibinfo {year} {2022})},\ \Eprint
  {http://arxiv.org/abs/2112.11588} {arXiv:2112.11588 [hep-ph]} \BibitemShut
  {NoStop}%
\bibitem [{\citenamefont {Kaplan}(1991)}]{Kaplan:1991dc}%
  \BibitemOpen
  \bibfield  {author} {\bibinfo {author} {\bibfnamefont {D.~B.}\ \bibnamefont
  {Kaplan}},\ }\href {\doibase 10.1016/S0550-3213(05)80021-5} {\bibfield
  {journal} {\bibinfo  {journal} {Nucl. Phys.}\ }\textbf {\bibinfo {volume}
  {B365}},\ \bibinfo {pages} {259} (\bibinfo {year} {1991})}\BibitemShut
  {NoStop}%
\bibitem [{\citenamefont {Khosa}\ and\ \citenamefont
  {Sanz}(2022)}]{Khosa:2021wsu}%
  \BibitemOpen
  \bibfield  {author} {\bibinfo {author} {\bibfnamefont {C.~K.}\ \bibnamefont
  {Khosa}}\ and\ \bibinfo {author} {\bibfnamefont {V.}~\bibnamefont {Sanz}},\
  }\href {\doibase 10.1155/2022/8970837} {\bibfield  {journal} {\bibinfo
  {journal} {Adv. High Energy Phys.}\ }\textbf {\bibinfo {volume} {2022}},\
  \bibinfo {pages} {8970837} (\bibinfo {year} {2022})},\ \Eprint
  {http://arxiv.org/abs/2102.13429} {arXiv:2102.13429 [hep-ph]} \BibitemShut
  {NoStop}%
\bibitem [{\citenamefont {Chung}(2021{\natexlab{b}})}]{Chung:2021ekz}%
  \BibitemOpen
  \bibfield  {author} {\bibinfo {author} {\bibfnamefont {Y.}~\bibnamefont
  {Chung}},\ }\href {\doibase 10.1103/PhysRevD.104.115027} {\bibfield
  {journal} {\bibinfo  {journal} {Phys. Rev. D}\ }\textbf {\bibinfo {volume}
  {104}},\ \bibinfo {pages} {115027} (\bibinfo {year} {2021}{\natexlab{b}})},\
  \Eprint {http://arxiv.org/abs/2108.08511} {arXiv:2108.08511 [hep-ph]}
  \BibitemShut {NoStop}%
\bibitem [{\citenamefont {Chung}(2021{\natexlab{c}})}]{Chung:2021xhd}%
  \BibitemOpen
  \bibfield  {author} {\bibinfo {author} {\bibfnamefont {Y.}~\bibnamefont
  {Chung}},\ }\href@noop {} {\  (\bibinfo {year} {2021}{\natexlab{c}})},\
  \Eprint {http://arxiv.org/abs/2110.03125} {arXiv:2110.03125 [hep-ph]}
  \BibitemShut {NoStop}%
\bibitem [{CMS(2022{\natexlab{a}})}]{CMS:2022ncp}%
  \BibitemOpen
  \href@noop {} {\bibfield  {journal} {\bibinfo  {journal} {CMS,
  CMS-EXO-21-009}\ } (\bibinfo {year} {2022}{\natexlab{a}})},\ \Eprint
  {http://arxiv.org/abs/2212.12604} {arXiv:2212.12604 [hep-ex]} \BibitemShut
  {NoStop}%
\bibitem [{ATL(2021{\natexlab{a}})}]{ATLAS:2021bjk}%
  \BibitemOpen
  \href@noop {} {\bibfield  {journal} {\bibinfo  {journal} {ATLAS,
  ATLAS-CONF-2021-025}\ } (\bibinfo {year} {2021}{\natexlab{a}})}\BibitemShut
  {NoStop}%
\bibitem [{CMS(2023)}]{CMS:2023ldh}%
  \BibitemOpen
  \href@noop {} {\bibfield  {journal} {\bibinfo  {journal} {CMS,
  CMS-PAS-B2G-20-012}\ } (\bibinfo {year} {2023})}\BibitemShut {NoStop}%
\bibitem [{ATL(2021{\natexlab{b}})}]{ATLAS:2021drn}%
  \BibitemOpen
  \href@noop {} {\bibfield  {journal} {\bibinfo  {journal} {ATLAS,
  ATLAS-CONF-2021-043}\ } (\bibinfo {year} {2021}{\natexlab{b}})}\BibitemShut
  {NoStop}%
\bibitem [{CMS(2022{\natexlab{b}})}]{CMS:2022goy}%
  \BibitemOpen
  \href@noop {} {\bibfield  {journal} {\bibinfo  {journal} {CMS,
  CMS-HIG-21-001}\ } (\bibinfo {year} {2022}{\natexlab{b}})},\ \Eprint
  {http://arxiv.org/abs/2208.02717} {arXiv:2208.02717 [hep-ex]} \BibitemShut
  {NoStop}%
\bibitem [{\citenamefont {Aad}\ \emph {et~al.}(2020{\natexlab{a}})\citenamefont
  {Aad} \emph {et~al.}}]{ATLAS:2020zms}%
  \BibitemOpen
  \bibfield  {author} {\bibinfo {author} {\bibfnamefont {G.}~\bibnamefont
  {Aad}} \emph {et~al.} (\bibinfo {collaboration} {ATLAS}),\ }\href {\doibase
  10.1103/PhysRevLett.125.051801} {\bibfield  {journal} {\bibinfo  {journal}
  {Phys. Rev. Lett.}\ }\textbf {\bibinfo {volume} {125}},\ \bibinfo {pages}
  {051801} (\bibinfo {year} {2020}{\natexlab{a}})},\ \Eprint
  {http://arxiv.org/abs/2002.12223} {arXiv:2002.12223 [hep-ex]} \BibitemShut
  {NoStop}%
\bibitem [{\citenamefont {Aad}\ \emph {et~al.}(2020{\natexlab{b}})\citenamefont
  {Aad} \emph {et~al.}}]{ATLAS:2020lks}%
  \BibitemOpen
  \bibfield  {author} {\bibinfo {author} {\bibfnamefont {G.}~\bibnamefont
  {Aad}} \emph {et~al.} (\bibinfo {collaboration} {ATLAS}),\ }\href {\doibase
  10.1007/JHEP10(2020)061} {\bibfield  {journal} {\bibinfo  {journal} {JHEP}\
  }\textbf {\bibinfo {volume} {10}},\ \bibinfo {pages} {061} (\bibinfo {year}
  {2020}{\natexlab{b}})},\ \Eprint {http://arxiv.org/abs/2005.05138}
  {arXiv:2005.05138 [hep-ex]} \BibitemShut {NoStop}%
\bibitem [{CMS(2022{\natexlab{c}})}]{CMS:2022zoc}%
  \BibitemOpen
  \href@noop {} {\  (\bibinfo {year} {2022}{\natexlab{c}})},\ \Eprint
  {http://arxiv.org/abs/2205.01835} {arXiv:2205.01835 [hep-ex]} \BibitemShut
  {NoStop}%
\bibitem [{\citenamefont {Aad}\ \emph {et~al.}(2020{\natexlab{c}})\citenamefont
  {Aad} \emph {et~al.}}]{ATLAS:2019fgd}%
  \BibitemOpen
  \bibfield  {author} {\bibinfo {author} {\bibfnamefont {G.}~\bibnamefont
  {Aad}} \emph {et~al.} (\bibinfo {collaboration} {ATLAS}),\ }\href {\doibase
  10.1007/JHEP03(2020)145} {\bibfield  {journal} {\bibinfo  {journal} {JHEP}\
  }\textbf {\bibinfo {volume} {03}},\ \bibinfo {pages} {145} (\bibinfo {year}
  {2020}{\natexlab{c}})},\ \Eprint {http://arxiv.org/abs/1910.08447}
  {arXiv:1910.08447 [hep-ex]} \BibitemShut {NoStop}%
\bibitem [{\citenamefont {Sirunyan}\ \emph {et~al.}(2019)\citenamefont
  {Sirunyan} \emph {et~al.}}]{CMS:2019nrx}%
  \BibitemOpen
  \bibfield  {author} {\bibinfo {author} {\bibfnamefont {A.~M.}\ \bibnamefont
  {Sirunyan}} \emph {et~al.} (\bibinfo {collaboration} {CMS}),\ }\href
  {\doibase 10.1103/PhysRevD.100.072002} {\bibfield  {journal} {\bibinfo
  {journal} {Phys. Rev. D}\ }\textbf {\bibinfo {volume} {100}},\ \bibinfo
  {pages} {072002} (\bibinfo {year} {2019})},\ \Eprint
  {http://arxiv.org/abs/1907.03729} {arXiv:1907.03729 [hep-ex]} \BibitemShut
  {NoStop}%
\bibitem [{\citenamefont {Aad}\ \emph {et~al.}(2022)\citenamefont {Aad} \emph
  {et~al.}}]{ATLAS:2022hwc}%
  \BibitemOpen
  \bibfield  {author} {\bibinfo {author} {\bibfnamefont {G.}~\bibnamefont
  {Aad}} \emph {et~al.} (\bibinfo {collaboration} {ATLAS}),\ }\href {\doibase
  10.1103/PhysRevD.105.092002} {\bibfield  {journal} {\bibinfo  {journal}
  {Phys. Rev. D}\ }\textbf {\bibinfo {volume} {105}},\ \bibinfo {pages}
  {092002} (\bibinfo {year} {2022})},\ \Eprint
  {http://arxiv.org/abs/2202.07288} {arXiv:2202.07288 [hep-ex]} \BibitemShut
  {NoStop}%
\bibitem [{\citenamefont {CMS}(2021)}]{CMS:2021qvd}%
  \BibitemOpen
  \bibfield  {author} {\bibinfo {author} {\bibnamefont {CMS}},\ }\href@noop {}
  {\bibfield  {journal} {\bibinfo  {journal} {CMS-PAS-B2G-20-004}\ } (\bibinfo
  {year} {2021})}\BibitemShut {NoStop}%
\bibitem [{\citenamefont {ATLAS}(2018)}]{ATLAS:2018crf}%
  \BibitemOpen
  \bibfield  {author} {\bibinfo {author} {\bibnamefont {ATLAS}},\ }\href@noop
  {} {\bibfield  {journal} {\bibinfo  {journal} {ATL-PHYS-PUB-2018-028}\ }
  (\bibinfo {year} {2018})}\BibitemShut {NoStop}%
\bibitem [{\citenamefont {Cid~Vidal}\ \emph {et~al.}(2019)\citenamefont
  {Cid~Vidal} \emph {et~al.}}]{CidVidal:2018eel}%
  \BibitemOpen
  \bibfield  {author} {\bibinfo {author} {\bibfnamefont {X.}~\bibnamefont
  {Cid~Vidal}} \emph {et~al.},\ }\href {\doibase 10.23731/CYRM-2019-007.585}
  {\bibfield  {journal} {\bibinfo  {journal} {CERN Yellow Rep. Monogr.}\
  }\textbf {\bibinfo {volume} {7}},\ \bibinfo {pages} {585} (\bibinfo {year}
  {2019})},\ \Eprint {http://arxiv.org/abs/1812.07831} {arXiv:1812.07831
  [hep-ph]} \BibitemShut {NoStop}%
\bibitem [{\citenamefont {Barbieri}\ \emph {et~al.}(2006)\citenamefont
  {Barbieri}, \citenamefont {Hall},\ and\ \citenamefont
  {Rychkov}}]{Barbieri:2006dq}%
  \BibitemOpen
  \bibfield  {author} {\bibinfo {author} {\bibfnamefont {R.}~\bibnamefont
  {Barbieri}}, \bibinfo {author} {\bibfnamefont {L.~J.}\ \bibnamefont {Hall}},
  \ and\ \bibinfo {author} {\bibfnamefont {V.~S.}\ \bibnamefont {Rychkov}},\
  }\href {\doibase 10.1103/PhysRevD.74.015007} {\bibfield  {journal} {\bibinfo
  {journal} {Phys. Rev. D}\ }\textbf {\bibinfo {volume} {74}},\ \bibinfo
  {pages} {015007} (\bibinfo {year} {2006})},\ \Eprint
  {http://arxiv.org/abs/hep-ph/0603188} {arXiv:hep-ph/0603188} \BibitemShut
  {NoStop}%
\bibitem [{\citenamefont {Lopez~Honorez}\ \emph {et~al.}(2007)\citenamefont
  {Lopez~Honorez}, \citenamefont {Nezri}, \citenamefont {Oliver},\ and\
  \citenamefont {Tytgat}}]{LopezHonorez:2006gr}%
  \BibitemOpen
  \bibfield  {author} {\bibinfo {author} {\bibfnamefont {L.}~\bibnamefont
  {Lopez~Honorez}}, \bibinfo {author} {\bibfnamefont {E.}~\bibnamefont
  {Nezri}}, \bibinfo {author} {\bibfnamefont {J.~F.}\ \bibnamefont {Oliver}}, \
  and\ \bibinfo {author} {\bibfnamefont {M.~H.~G.}\ \bibnamefont {Tytgat}},\
  }\href {\doibase 10.1088/1475-7516/2007/02/028} {\bibfield  {journal}
  {\bibinfo  {journal} {JCAP}\ }\textbf {\bibinfo {volume} {02}},\ \bibinfo
  {pages} {028} (\bibinfo {year} {2007})},\ \Eprint
  {http://arxiv.org/abs/hep-ph/0612275} {arXiv:hep-ph/0612275} \BibitemShut
  {NoStop}%
\bibitem [{\citenamefont {Chowdhury}\ \emph {et~al.}(2012)\citenamefont
  {Chowdhury}, \citenamefont {Nemevsek}, \citenamefont {Senjanovic},\ and\
  \citenamefont {Zhang}}]{Chowdhury:2011ga}%
  \BibitemOpen
  \bibfield  {author} {\bibinfo {author} {\bibfnamefont {T.~A.}\ \bibnamefont
  {Chowdhury}}, \bibinfo {author} {\bibfnamefont {M.}~\bibnamefont {Nemevsek}},
  \bibinfo {author} {\bibfnamefont {G.}~\bibnamefont {Senjanovic}}, \ and\
  \bibinfo {author} {\bibfnamefont {Y.}~\bibnamefont {Zhang}},\ }\href
  {\doibase 10.1088/1475-7516/2012/02/029} {\bibfield  {journal} {\bibinfo
  {journal} {JCAP}\ }\textbf {\bibinfo {volume} {02}},\ \bibinfo {pages} {029}
  (\bibinfo {year} {2012})},\ \Eprint {http://arxiv.org/abs/1110.5334}
  {arXiv:1110.5334 [hep-ph]} \BibitemShut {NoStop}%
\bibitem [{\citenamefont {Borah}\ and\ \citenamefont
  {Cline}(2012)}]{Borah:2012pu}%
  \BibitemOpen
  \bibfield  {author} {\bibinfo {author} {\bibfnamefont {D.}~\bibnamefont
  {Borah}}\ and\ \bibinfo {author} {\bibfnamefont {J.~M.}\ \bibnamefont
  {Cline}},\ }\href {\doibase 10.1103/PhysRevD.86.055001} {\bibfield  {journal}
  {\bibinfo  {journal} {Phys. Rev. D}\ }\textbf {\bibinfo {volume} {86}},\
  \bibinfo {pages} {055001} (\bibinfo {year} {2012})},\ \Eprint
  {http://arxiv.org/abs/1204.4722} {arXiv:1204.4722 [hep-ph]} \BibitemShut
  {NoStop}%
\bibitem [{\citenamefont {Cline}\ and\ \citenamefont
  {Kainulainen}(2013)}]{Cline:2013bln}%
  \BibitemOpen
  \bibfield  {author} {\bibinfo {author} {\bibfnamefont {J.~M.}\ \bibnamefont
  {Cline}}\ and\ \bibinfo {author} {\bibfnamefont {K.}~\bibnamefont
  {Kainulainen}},\ }\href {\doibase 10.1103/PhysRevD.87.071701} {\bibfield
  {journal} {\bibinfo  {journal} {Phys. Rev. D}\ }\textbf {\bibinfo {volume}
  {87}},\ \bibinfo {pages} {071701} (\bibinfo {year} {2013})},\ \Eprint
  {http://arxiv.org/abs/1302.2614} {arXiv:1302.2614 [hep-ph]} \BibitemShut
  {NoStop}%
\bibitem [{\citenamefont {Blinov}\ \emph {et~al.}(2015)\citenamefont {Blinov},
  \citenamefont {Profumo},\ and\ \citenamefont {Stefaniak}}]{Blinov:2015vma}%
  \BibitemOpen
  \bibfield  {author} {\bibinfo {author} {\bibfnamefont {N.}~\bibnamefont
  {Blinov}}, \bibinfo {author} {\bibfnamefont {S.}~\bibnamefont {Profumo}}, \
  and\ \bibinfo {author} {\bibfnamefont {T.}~\bibnamefont {Stefaniak}},\ }\href
  {\doibase 10.1088/1475-7516/2015/07/028} {\bibfield  {journal} {\bibinfo
  {journal} {JCAP}\ }\textbf {\bibinfo {volume} {07}},\ \bibinfo {pages} {028}
  (\bibinfo {year} {2015})},\ \Eprint {http://arxiv.org/abs/1504.05949}
  {arXiv:1504.05949 [hep-ph]} \BibitemShut {NoStop}%
\bibitem [{\citenamefont {Ilnicka}\ \emph {et~al.}(2016)\citenamefont
  {Ilnicka}, \citenamefont {Krawczyk},\ and\ \citenamefont
  {Robens}}]{Ilnicka:2015jba}%
  \BibitemOpen
  \bibfield  {author} {\bibinfo {author} {\bibfnamefont {A.}~\bibnamefont
  {Ilnicka}}, \bibinfo {author} {\bibfnamefont {M.}~\bibnamefont {Krawczyk}}, \
  and\ \bibinfo {author} {\bibfnamefont {T.}~\bibnamefont {Robens}},\ }\href
  {\doibase 10.1103/PhysRevD.93.055026} {\bibfield  {journal} {\bibinfo
  {journal} {Phys. Rev. D}\ }\textbf {\bibinfo {volume} {93}},\ \bibinfo
  {pages} {055026} (\bibinfo {year} {2016})},\ \Eprint
  {http://arxiv.org/abs/1508.01671} {arXiv:1508.01671 [hep-ph]} \BibitemShut
  {NoStop}%
\bibitem [{\citenamefont {Belyaev}\ \emph {et~al.}(2018)\citenamefont
  {Belyaev}, \citenamefont {Cacciapaglia}, \citenamefont {Ivanov},
  \citenamefont {Rojas-Abatte},\ and\ \citenamefont
  {Thomas}}]{Belyaev:2016lok}%
  \BibitemOpen
  \bibfield  {author} {\bibinfo {author} {\bibfnamefont {A.}~\bibnamefont
  {Belyaev}}, \bibinfo {author} {\bibfnamefont {G.}~\bibnamefont
  {Cacciapaglia}}, \bibinfo {author} {\bibfnamefont {I.~P.}\ \bibnamefont
  {Ivanov}}, \bibinfo {author} {\bibfnamefont {F.}~\bibnamefont
  {Rojas-Abatte}}, \ and\ \bibinfo {author} {\bibfnamefont {M.}~\bibnamefont
  {Thomas}},\ }\href {\doibase 10.1103/PhysRevD.97.035011} {\bibfield
  {journal} {\bibinfo  {journal} {Phys. Rev. D}\ }\textbf {\bibinfo {volume}
  {97}},\ \bibinfo {pages} {035011} (\bibinfo {year} {2018})},\ \Eprint
  {http://arxiv.org/abs/1612.00511} {arXiv:1612.00511 [hep-ph]} \BibitemShut
  {NoStop}%
\bibitem [{\citenamefont {Fabian}\ \emph {et~al.}(2021)\citenamefont {Fabian},
  \citenamefont {Goertz},\ and\ \citenamefont {Jiang}}]{Fabian:2020hny}%
  \BibitemOpen
  \bibfield  {author} {\bibinfo {author} {\bibfnamefont {S.}~\bibnamefont
  {Fabian}}, \bibinfo {author} {\bibfnamefont {F.}~\bibnamefont {Goertz}}, \
  and\ \bibinfo {author} {\bibfnamefont {Y.}~\bibnamefont {Jiang}},\ }\href
  {\doibase 10.1088/1475-7516/2021/09/011} {\bibfield  {journal} {\bibinfo
  {journal} {JCAP}\ }\textbf {\bibinfo {volume} {09}},\ \bibinfo {pages} {011}
  (\bibinfo {year} {2021})},\ \Eprint {http://arxiv.org/abs/2012.12847}
  {arXiv:2012.12847 [hep-ph]} \BibitemShut {NoStop}%
\bibitem [{\citenamefont {Astros}\ \emph {et~al.}(2023)\citenamefont {Astros},
  \citenamefont {Fabian},\ and\ \citenamefont {Goertz}}]{Astros:2023gda}%
  \BibitemOpen
  \bibfield  {author} {\bibinfo {author} {\bibfnamefont {M.~D.}\ \bibnamefont
  {Astros}}, \bibinfo {author} {\bibfnamefont {S.}~\bibnamefont {Fabian}}, \
  and\ \bibinfo {author} {\bibfnamefont {F.}~\bibnamefont {Goertz}},\
  }\href@noop {} {\  (\bibinfo {year} {2023})},\ \Eprint
  {http://arxiv.org/abs/2307.01270} {arXiv:2307.01270 [hep-ph]} \BibitemShut
  {NoStop}%
\bibitem [{\citenamefont {Mammen~Abraham}\ \emph {et~al.}(2022)\citenamefont
  {Mammen~Abraham}, \citenamefont {Gon\c{c}alves}, \citenamefont {Han},
  \citenamefont {Leung},\ and\ \citenamefont {Qin}}]{MammenAbraham:2021ssc}%
  \BibitemOpen
  \bibfield  {author} {\bibinfo {author} {\bibfnamefont {R.}~\bibnamefont
  {Mammen~Abraham}}, \bibinfo {author} {\bibfnamefont {D.}~\bibnamefont
  {Gon\c{c}alves}}, \bibinfo {author} {\bibfnamefont {T.}~\bibnamefont {Han}},
  \bibinfo {author} {\bibfnamefont {S.~C.~I.}\ \bibnamefont {Leung}}, \ and\
  \bibinfo {author} {\bibfnamefont {H.}~\bibnamefont {Qin}},\ }\href {\doibase
  10.1016/j.physletb.2021.136839} {\bibfield  {journal} {\bibinfo  {journal}
  {Phys. Lett. B}\ }\textbf {\bibinfo {volume} {825}},\ \bibinfo {pages}
  {136839} (\bibinfo {year} {2022})},\ \Eprint
  {http://arxiv.org/abs/2106.00018} {arXiv:2106.00018 [hep-ph]} \BibitemShut
  {NoStop}%
\bibitem [{\citenamefont {Bittar}\ and\ \citenamefont
  {Burdman}(2022)}]{Bittar:2022wgb}%
  \BibitemOpen
  \bibfield  {author} {\bibinfo {author} {\bibfnamefont {P.}~\bibnamefont
  {Bittar}}\ and\ \bibinfo {author} {\bibfnamefont {G.}~\bibnamefont
  {Burdman}},\ }\href {\doibase 10.1007/JHEP10(2022)004} {\bibfield  {journal}
  {\bibinfo  {journal} {JHEP}\ }\textbf {\bibinfo {volume} {10}},\ \bibinfo
  {pages} {004} (\bibinfo {year} {2022})},\ \Eprint
  {http://arxiv.org/abs/2204.07094} {arXiv:2204.07094 [hep-ph]} \BibitemShut
  {NoStop}%
\bibitem [{\citenamefont {Sirunyan}\ \emph {et~al.}(2020)\citenamefont
  {Sirunyan} \emph {et~al.}}]{CMS:2019jul}%
  \BibitemOpen
  \bibfield  {author} {\bibinfo {author} {\bibfnamefont {A.~M.}\ \bibnamefont
  {Sirunyan}} \emph {et~al.} (\bibinfo {collaboration} {CMS}),\ }\href
  {\doibase 10.1016/j.physletb.2020.135263} {\bibfield  {journal} {\bibinfo
  {journal} {Phys. Lett. B}\ }\textbf {\bibinfo {volume} {803}},\ \bibinfo
  {pages} {135263} (\bibinfo {year} {2020})},\ \Eprint
  {http://arxiv.org/abs/1909.09193} {arXiv:1909.09193 [hep-ex]} \BibitemShut
  {NoStop}%
\bibitem [{\citenamefont {Defranchis}\ \emph {et~al.}(2022)\citenamefont
  {Defranchis}, \citenamefont {Kieseler}, \citenamefont {Lipka},\ and\
  \citenamefont {Mazzitelli}}]{Defranchis:2022nqb}%
  \BibitemOpen
  \bibfield  {author} {\bibinfo {author} {\bibfnamefont {M.~M.}\ \bibnamefont
  {Defranchis}}, \bibinfo {author} {\bibfnamefont {J.}~\bibnamefont
  {Kieseler}}, \bibinfo {author} {\bibfnamefont {K.}~\bibnamefont {Lipka}}, \
  and\ \bibinfo {author} {\bibfnamefont {J.}~\bibnamefont {Mazzitelli}},\
  }\href@noop {} {\  (\bibinfo {year} {2022})},\ \Eprint
  {http://arxiv.org/abs/2208.11399} {arXiv:2208.11399 [hep-ph]} \BibitemShut
  {NoStop}%
\bibitem [{\citenamefont {van Beekveld}\ \emph {et~al.}(2023)\citenamefont {van
  Beekveld}, \citenamefont {Kulesza},\ and\ \citenamefont
  {Valero}}]{vanBeekveld:2022hty}%
  \BibitemOpen
  \bibfield  {author} {\bibinfo {author} {\bibfnamefont {M.}~\bibnamefont {van
  Beekveld}}, \bibinfo {author} {\bibfnamefont {A.}~\bibnamefont {Kulesza}}, \
  and\ \bibinfo {author} {\bibfnamefont {L.~M.}\ \bibnamefont {Valero}},\
  }\href {\doibase 10.1103/PhysRevLett.131.211901} {\bibfield  {journal}
  {\bibinfo  {journal} {Phys. Rev. Lett.}\ }\textbf {\bibinfo {volume} {131}},\
  \bibinfo {pages} {211901} (\bibinfo {year} {2023})},\ \Eprint
  {http://arxiv.org/abs/2212.03259} {arXiv:2212.03259 [hep-ph]} \BibitemShut
  {NoStop}%
\bibitem [{\citenamefont {Aad}\ \emph {et~al.}(2023)\citenamefont {Aad} \emph
  {et~al.}}]{ATLAS:2023ajo}%
  \BibitemOpen
  \bibfield  {author} {\bibinfo {author} {\bibfnamefont {G.}~\bibnamefont
  {Aad}} \emph {et~al.} (\bibinfo {collaboration} {ATLAS}),\ }\href@noop {} {\
  (\bibinfo {year} {2023})},\ \Eprint {http://arxiv.org/abs/2303.15061}
  {arXiv:2303.15061 [hep-ex]} \BibitemShut {NoStop}%
\bibitem [{\citenamefont {Hayrapetyan}\ \emph {et~al.}(2023)\citenamefont
  {Hayrapetyan} \emph {et~al.}}]{CMS:2023ftu}%
  \BibitemOpen
  \bibfield  {author} {\bibinfo {author} {\bibfnamefont {A.}~\bibnamefont
  {Hayrapetyan}} \emph {et~al.} (\bibinfo {collaboration} {CMS}),\ }\href@noop
  {} {\  (\bibinfo {year} {2023})},\ \Eprint {http://arxiv.org/abs/2305.13439}
  {arXiv:2305.13439 [hep-ex]} \BibitemShut {NoStop}%
\bibitem [{\citenamefont {Darm\'e}\ \emph {et~al.}(2021)\citenamefont
  {Darm\'e}, \citenamefont {Fuks},\ and\ \citenamefont
  {Maltoni}}]{Darme:2021gtt}%
  \BibitemOpen
  \bibfield  {author} {\bibinfo {author} {\bibfnamefont {L.}~\bibnamefont
  {Darm\'e}}, \bibinfo {author} {\bibfnamefont {B.}~\bibnamefont {Fuks}}, \
  and\ \bibinfo {author} {\bibfnamefont {F.}~\bibnamefont {Maltoni}},\ }\href
  {\doibase 10.1007/JHEP09(2021)143} {\bibfield  {journal} {\bibinfo  {journal}
  {JHEP}\ }\textbf {\bibinfo {volume} {09}},\ \bibinfo {pages} {143} (\bibinfo
  {year} {2021})},\ \Eprint {http://arxiv.org/abs/2104.09512} {arXiv:2104.09512
  [hep-ph]} \BibitemShut {NoStop}%
\bibitem [{\citenamefont {Banelli}\ \emph {et~al.}(2021)\citenamefont
  {Banelli}, \citenamefont {Salvioni}, \citenamefont {Serra}, \citenamefont
  {Theil},\ and\ \citenamefont {Weiler}}]{Banelli:2020iau}%
  \BibitemOpen
  \bibfield  {author} {\bibinfo {author} {\bibfnamefont {G.}~\bibnamefont
  {Banelli}}, \bibinfo {author} {\bibfnamefont {E.}~\bibnamefont {Salvioni}},
  \bibinfo {author} {\bibfnamefont {J.}~\bibnamefont {Serra}}, \bibinfo
  {author} {\bibfnamefont {T.}~\bibnamefont {Theil}}, \ and\ \bibinfo {author}
  {\bibfnamefont {A.}~\bibnamefont {Weiler}},\ }\href {\doibase
  10.1007/JHEP02(2021)043} {\bibfield  {journal} {\bibinfo  {journal} {JHEP}\
  }\textbf {\bibinfo {volume} {02}},\ \bibinfo {pages} {043} (\bibinfo {year}
  {2021})},\ \Eprint {http://arxiv.org/abs/2010.05915} {arXiv:2010.05915
  [hep-ph]} \BibitemShut {NoStop}%
\bibitem [{\citenamefont {Blekman}\ \emph {et~al.}(2022)\citenamefont
  {Blekman}, \citenamefont {D\'eliot}, \citenamefont {Dutta},\ and\
  \citenamefont {Usai}}]{Blekman:2022jag}%
  \BibitemOpen
  \bibfield  {author} {\bibinfo {author} {\bibfnamefont {F.}~\bibnamefont
  {Blekman}}, \bibinfo {author} {\bibfnamefont {F.}~\bibnamefont {D\'eliot}},
  \bibinfo {author} {\bibfnamefont {V.}~\bibnamefont {Dutta}}, \ and\ \bibinfo
  {author} {\bibfnamefont {E.}~\bibnamefont {Usai}},\ }\href {\doibase
  10.3390/universe8120638} {\bibfield  {journal} {\bibinfo  {journal}
  {Universe}\ }\textbf {\bibinfo {volume} {8}},\ \bibinfo {pages} {638}
  (\bibinfo {year} {2022})},\ \Eprint {http://arxiv.org/abs/2208.04085}
  {arXiv:2208.04085 [hep-ex]} \BibitemShut {NoStop}%
\bibitem [{\citenamefont {Di~Luzio}\ \emph {et~al.}(2018)\citenamefont
  {Di~Luzio}, \citenamefont {Kirk},\ and\ \citenamefont
  {Lenz}}]{DiLuzio:2017fdq}%
  \BibitemOpen
  \bibfield  {author} {\bibinfo {author} {\bibfnamefont {L.}~\bibnamefont
  {Di~Luzio}}, \bibinfo {author} {\bibfnamefont {M.}~\bibnamefont {Kirk}}, \
  and\ \bibinfo {author} {\bibfnamefont {A.}~\bibnamefont {Lenz}},\ }\href
  {\doibase 10.1103/PhysRevD.97.095035} {\bibfield  {journal} {\bibinfo
  {journal} {Phys. Rev. D}\ }\textbf {\bibinfo {volume} {97}},\ \bibinfo
  {pages} {095035} (\bibinfo {year} {2018})},\ \Eprint
  {http://arxiv.org/abs/1712.06572} {arXiv:1712.06572 [hep-ph]} \BibitemShut
  {NoStop}%
\bibitem [{\citenamefont {Allanach}\ \emph {et~al.}(2019)\citenamefont
  {Allanach}, \citenamefont {Butterworth},\ and\ \citenamefont
  {Corbett}}]{Allanach:2019mfl}%
  \BibitemOpen
  \bibfield  {author} {\bibinfo {author} {\bibfnamefont {B.~C.}\ \bibnamefont
  {Allanach}}, \bibinfo {author} {\bibfnamefont {J.~M.}\ \bibnamefont
  {Butterworth}}, \ and\ \bibinfo {author} {\bibfnamefont {T.}~\bibnamefont
  {Corbett}},\ }\href {\doibase 10.1007/JHEP08(2019)106} {\bibfield  {journal}
  {\bibinfo  {journal} {JHEP}\ }\textbf {\bibinfo {volume} {08}},\ \bibinfo
  {pages} {106} (\bibinfo {year} {2019})},\ \Eprint
  {http://arxiv.org/abs/1904.10954} {arXiv:1904.10954 [hep-ph]} \BibitemShut
  {NoStop}%
\bibitem [{\citenamefont {Peskin}\ and\ \citenamefont
  {Takeuchi}(1990)}]{Peskin:1990zt}%
  \BibitemOpen
  \bibfield  {author} {\bibinfo {author} {\bibfnamefont {M.~E.}\ \bibnamefont
  {Peskin}}\ and\ \bibinfo {author} {\bibfnamefont {T.}~\bibnamefont
  {Takeuchi}},\ }\href {\doibase 10.1103/PhysRevLett.65.964} {\bibfield
  {journal} {\bibinfo  {journal} {Phys. Rev. Lett.}\ }\textbf {\bibinfo
  {volume} {65}},\ \bibinfo {pages} {964} (\bibinfo {year} {1990})}\BibitemShut
  {NoStop}%
\bibitem [{\citenamefont {Peskin}\ and\ \citenamefont
  {Takeuchi}(1992)}]{Peskin:1991sw}%
  \BibitemOpen
  \bibfield  {author} {\bibinfo {author} {\bibfnamefont {M.~E.}\ \bibnamefont
  {Peskin}}\ and\ \bibinfo {author} {\bibfnamefont {T.}~\bibnamefont
  {Takeuchi}},\ }\href {\doibase 10.1103/PhysRevD.46.381} {\bibfield  {journal}
  {\bibinfo  {journal} {Phys. Rev. D}\ }\textbf {\bibinfo {volume} {46}},\
  \bibinfo {pages} {381} (\bibinfo {year} {1992})}\BibitemShut {NoStop}%
\bibitem [{\citenamefont {Workman}\ \emph {et~al.}(2022)\citenamefont {Workman}
  \emph {et~al.}}]{ParticleDataGroup:2022pth}%
  \BibitemOpen
  \bibfield  {author} {\bibinfo {author} {\bibfnamefont {R.~L.}\ \bibnamefont
  {Workman}} \emph {et~al.} (\bibinfo {collaboration} {Particle Data Group}),\
  }\href {\doibase 10.1093/ptep/ptac097} {\bibfield  {journal} {\bibinfo
  {journal} {PTEP}\ }\textbf {\bibinfo {volume} {2022}},\ \bibinfo {pages}
  {083C01} (\bibinfo {year} {2022})}\BibitemShut {NoStop}%
\bibitem [{\citenamefont {Barbieri}\ \emph {et~al.}(2007)\citenamefont
  {Barbieri}, \citenamefont {Bellazzini}, \citenamefont {Rychkov},\ and\
  \citenamefont {Varagnolo}}]{Barbieri:2007bh}%
  \BibitemOpen
  \bibfield  {author} {\bibinfo {author} {\bibfnamefont {R.}~\bibnamefont
  {Barbieri}}, \bibinfo {author} {\bibfnamefont {B.}~\bibnamefont
  {Bellazzini}}, \bibinfo {author} {\bibfnamefont {V.~S.}\ \bibnamefont
  {Rychkov}}, \ and\ \bibinfo {author} {\bibfnamefont {A.}~\bibnamefont
  {Varagnolo}},\ }\href {\doibase 10.1103/PhysRevD.76.115008} {\bibfield
  {journal} {\bibinfo  {journal} {Phys. Rev. D}\ }\textbf {\bibinfo {volume}
  {76}},\ \bibinfo {pages} {115008} (\bibinfo {year} {2007})},\ \Eprint
  {http://arxiv.org/abs/0706.0432} {arXiv:0706.0432 [hep-ph]} \BibitemShut
  {NoStop}%
\bibitem [{\citenamefont {Grojean}\ \emph {et~al.}(2013)\citenamefont
  {Grojean}, \citenamefont {Matsedonskyi},\ and\ \citenamefont
  {Panico}}]{Grojean:2013qca}%
  \BibitemOpen
  \bibfield  {author} {\bibinfo {author} {\bibfnamefont {C.}~\bibnamefont
  {Grojean}}, \bibinfo {author} {\bibfnamefont {O.}~\bibnamefont
  {Matsedonskyi}}, \ and\ \bibinfo {author} {\bibfnamefont {G.}~\bibnamefont
  {Panico}},\ }\href {\doibase 10.1007/JHEP10(2013)160} {\bibfield  {journal}
  {\bibinfo  {journal} {JHEP}\ }\textbf {\bibinfo {volume} {10}},\ \bibinfo
  {pages} {160} (\bibinfo {year} {2013})},\ \Eprint
  {http://arxiv.org/abs/1306.4655} {arXiv:1306.4655 [hep-ph]} \BibitemShut
  {NoStop}%
\bibitem [{\citenamefont {Csaki}\ \emph
  {et~al.}(2003{\natexlab{a}})\citenamefont {Csaki}, \citenamefont {Hubisz},
  \citenamefont {Kribs}, \citenamefont {Meade},\ and\ \citenamefont
  {Terning}}]{Csaki:2002qg}%
  \BibitemOpen
  \bibfield  {author} {\bibinfo {author} {\bibfnamefont {C.}~\bibnamefont
  {Csaki}}, \bibinfo {author} {\bibfnamefont {J.}~\bibnamefont {Hubisz}},
  \bibinfo {author} {\bibfnamefont {G.~D.}\ \bibnamefont {Kribs}}, \bibinfo
  {author} {\bibfnamefont {P.}~\bibnamefont {Meade}}, \ and\ \bibinfo {author}
  {\bibfnamefont {J.}~\bibnamefont {Terning}},\ }\href {\doibase
  10.1103/PhysRevD.67.115002} {\bibfield  {journal} {\bibinfo  {journal} {Phys.
  Rev. D}\ }\textbf {\bibinfo {volume} {67}},\ \bibinfo {pages} {115002}
  (\bibinfo {year} {2003}{\natexlab{a}})},\ \Eprint
  {http://arxiv.org/abs/hep-ph/0211124} {arXiv:hep-ph/0211124} \BibitemShut
  {NoStop}%
\bibitem [{\citenamefont {Hewett}\ \emph {et~al.}(2003)\citenamefont {Hewett},
  \citenamefont {Petriello},\ and\ \citenamefont {Rizzo}}]{Hewett:2002px}%
  \BibitemOpen
  \bibfield  {author} {\bibinfo {author} {\bibfnamefont {J.~L.}\ \bibnamefont
  {Hewett}}, \bibinfo {author} {\bibfnamefont {F.~J.}\ \bibnamefont
  {Petriello}}, \ and\ \bibinfo {author} {\bibfnamefont {T.~G.}\ \bibnamefont
  {Rizzo}},\ }\href {\doibase 10.1088/1126-6708/2003/10/062} {\bibfield
  {journal} {\bibinfo  {journal} {JHEP}\ }\textbf {\bibinfo {volume} {10}},\
  \bibinfo {pages} {062} (\bibinfo {year} {2003})},\ \Eprint
  {http://arxiv.org/abs/hep-ph/0211218} {arXiv:hep-ph/0211218} \BibitemShut
  {NoStop}%
\bibitem [{\citenamefont {Csaki}\ \emph
  {et~al.}(2003{\natexlab{b}})\citenamefont {Csaki}, \citenamefont {Hubisz},
  \citenamefont {Kribs}, \citenamefont {Meade},\ and\ \citenamefont
  {Terning}}]{Csaki:2003si}%
  \BibitemOpen
  \bibfield  {author} {\bibinfo {author} {\bibfnamefont {C.}~\bibnamefont
  {Csaki}}, \bibinfo {author} {\bibfnamefont {J.}~\bibnamefont {Hubisz}},
  \bibinfo {author} {\bibfnamefont {G.~D.}\ \bibnamefont {Kribs}}, \bibinfo
  {author} {\bibfnamefont {P.}~\bibnamefont {Meade}}, \ and\ \bibinfo {author}
  {\bibfnamefont {J.}~\bibnamefont {Terning}},\ }\href {\doibase
  10.1103/PhysRevD.68.035009} {\bibfield  {journal} {\bibinfo  {journal} {Phys.
  Rev. D}\ }\textbf {\bibinfo {volume} {68}},\ \bibinfo {pages} {035009}
  (\bibinfo {year} {2003}{\natexlab{b}})},\ \Eprint
  {http://arxiv.org/abs/hep-ph/0303236} {arXiv:hep-ph/0303236} \BibitemShut
  {NoStop}%
\bibitem [{\citenamefont {Gregoire}\ \emph {et~al.}(2004)\citenamefont
  {Gregoire}, \citenamefont {Tucker-Smith},\ and\ \citenamefont
  {Wacker}}]{Gregoire:2003kr}%
  \BibitemOpen
  \bibfield  {author} {\bibinfo {author} {\bibfnamefont {T.}~\bibnamefont
  {Gregoire}}, \bibinfo {author} {\bibfnamefont {D.}~\bibnamefont
  {Tucker-Smith}}, \ and\ \bibinfo {author} {\bibfnamefont {J.~G.}\
  \bibnamefont {Wacker}},\ }\href {\doibase 10.1103/PhysRevD.69.115008}
  {\bibfield  {journal} {\bibinfo  {journal} {Phys. Rev. D}\ }\textbf {\bibinfo
  {volume} {69}},\ \bibinfo {pages} {115008} (\bibinfo {year} {2004})},\
  \Eprint {http://arxiv.org/abs/hep-ph/0305275} {arXiv:hep-ph/0305275}
  \BibitemShut {NoStop}%
\bibitem [{\citenamefont {Marandella}\ \emph {et~al.}(2005)\citenamefont
  {Marandella}, \citenamefont {Schappacher},\ and\ \citenamefont
  {Strumia}}]{Marandella:2005wd}%
  \BibitemOpen
  \bibfield  {author} {\bibinfo {author} {\bibfnamefont {G.}~\bibnamefont
  {Marandella}}, \bibinfo {author} {\bibfnamefont {C.}~\bibnamefont
  {Schappacher}}, \ and\ \bibinfo {author} {\bibfnamefont {A.}~\bibnamefont
  {Strumia}},\ }\href {\doibase 10.1103/PhysRevD.72.035014} {\bibfield
  {journal} {\bibinfo  {journal} {Phys. Rev. D}\ }\textbf {\bibinfo {volume}
  {72}},\ \bibinfo {pages} {035014} (\bibinfo {year} {2005})},\ \Eprint
  {http://arxiv.org/abs/hep-ph/0502096} {arXiv:hep-ph/0502096} \BibitemShut
  {NoStop}%
\bibitem [{\citenamefont {Strumia}(2022)}]{Strumia:2022qkt}%
  \BibitemOpen
  \bibfield  {author} {\bibinfo {author} {\bibfnamefont {A.}~\bibnamefont
  {Strumia}},\ }\href {\doibase 10.1007/JHEP08(2022)248} {\bibfield  {journal}
  {\bibinfo  {journal} {JHEP}\ }\textbf {\bibinfo {volume} {08}},\ \bibinfo
  {pages} {248} (\bibinfo {year} {2022})},\ \Eprint
  {http://arxiv.org/abs/2204.04191} {arXiv:2204.04191 [hep-ph]} \BibitemShut
  {NoStop}%
\bibitem [{\citenamefont {Batell}\ \emph {et~al.}(2013)\citenamefont {Batell},
  \citenamefont {Gori},\ and\ \citenamefont {Wang}}]{Batell:2012ca}%
  \BibitemOpen
  \bibfield  {author} {\bibinfo {author} {\bibfnamefont {B.}~\bibnamefont
  {Batell}}, \bibinfo {author} {\bibfnamefont {S.}~\bibnamefont {Gori}}, \ and\
  \bibinfo {author} {\bibfnamefont {L.-T.}\ \bibnamefont {Wang}},\ }\href
  {\doibase 10.1007/JHEP01(2013)139} {\bibfield  {journal} {\bibinfo  {journal}
  {JHEP}\ }\textbf {\bibinfo {volume} {01}},\ \bibinfo {pages} {139} (\bibinfo
  {year} {2013})},\ \Eprint {http://arxiv.org/abs/1209.6382} {arXiv:1209.6382
  [hep-ph]} \BibitemShut {NoStop}%
\bibitem [{\citenamefont {Guadagnoli}\ and\ \citenamefont
  {Isidori}(2013)}]{Guadagnoli:2013mru}%
  \BibitemOpen
  \bibfield  {author} {\bibinfo {author} {\bibfnamefont {D.}~\bibnamefont
  {Guadagnoli}}\ and\ \bibinfo {author} {\bibfnamefont {G.}~\bibnamefont
  {Isidori}},\ }\href {\doibase 10.1016/j.physletb.2013.05.054} {\bibfield
  {journal} {\bibinfo  {journal} {Phys. Lett. B}\ }\textbf {\bibinfo {volume}
  {724}},\ \bibinfo {pages} {63} (\bibinfo {year} {2013})},\ \Eprint
  {http://arxiv.org/abs/1302.3909} {arXiv:1302.3909 [hep-ph]} \BibitemShut
  {NoStop}%
\bibitem [{\citenamefont {Chivukula}\ \emph {et~al.}(1992)\citenamefont
  {Chivukula}, \citenamefont {Selipsky},\ and\ \citenamefont
  {Simmons}}]{Chivukula:1992ap}%
  \BibitemOpen
  \bibfield  {author} {\bibinfo {author} {\bibfnamefont {R.~S.}\ \bibnamefont
  {Chivukula}}, \bibinfo {author} {\bibfnamefont {S.~B.}\ \bibnamefont
  {Selipsky}}, \ and\ \bibinfo {author} {\bibfnamefont {E.~H.}\ \bibnamefont
  {Simmons}},\ }\href {\doibase 10.1103/PhysRevLett.69.575} {\bibfield
  {journal} {\bibinfo  {journal} {Phys. Rev. Lett.}\ }\textbf {\bibinfo
  {volume} {69}},\ \bibinfo {pages} {575} (\bibinfo {year} {1992})},\ \Eprint
  {http://arxiv.org/abs/hep-ph/9204214} {arXiv:hep-ph/9204214} \BibitemShut
  {NoStop}%
\end{thebibliography}%

\end{document}